\documentclass[10pt, twocolumn,aps,showpacs,prb]{revtex4-1}

\usepackage{amsmath}
\usepackage{amssymb}
\usepackage{amsfonts, epsfig}
\usepackage{graphicx}
\usepackage{setspace}
\usepackage{calc}
\usepackage{floatflt}

\def\H{{\cal H}}

\def\ket#1{|#1\rangle }
\def\bra#1{\langle#1 | }

\def\non{\nonumber \\ }

\begin{document}

\title{Coherent control of correlated nanodevices:
       A hybrid time-dependent numerical
       renormalization-group approach to periodic
       switching}

\author{Eitan Eidelstein}
\author{Avraham Schiller}
\affiliation{Racah Institute of Physics, The Hebrew
             University, Jerusalem 91904, Israel}
\author{Fabian G\"uttge}
\author{Frithjof B.~Anders}
\affiliation{Lehrstuhl f\"ur Theoretische Physik II,
             Technische Universit\"at Dortmund, 44221
             Dortmund,Germany}
\date{January 15, 2012}

\begin{abstract}
The time-dependent numerical renormalization-group
approach (TD-NRG), originally devised for tracking
the real-time dynamics of quantum-impurity systems
following a single quantum quench, is extended to
multiple switching events. This generalization of
the TD-NRG encompasses the possibility of periodic
switching, allowing for coherent control of strongly
correlated systems by an external time-dependent
field. To this end, we have  embedded  the TD-NRG in a
hybrid framework that combines the outstanding
capabilities of the numerical renormalization group
to systematically construct the effective low-energy
Hamiltonian of the system with the prowess of
complementary approaches for calculating the
real-time dynamics derived from this Hamiltonian. We
demonstrate the power of our approach by hybridizing
the TD-NRG with the Chebyshev expansion technique
in order to investigate periodic switching in the
interacting resonant-level model. Although
the interacting model shares the same low-energy
fixed point as its noninteracting counterpart, we
surprisingly find the gradual emergence of damped
oscillations as the interaction strength is
increased. Focusing on a single quantum quench and
using a strong-coupling analysis, we reveal the
origin of these interaction-induced oscillations
and provide an analytical estimate for their
frequency. The latter agrees well with the
numerical results.
\end{abstract}


\maketitle

\section{Introduction}

The quantitative description of real-time
dynamics in strongly correlated systems is one
of the outstanding challenges of contemporary
condensed-matter physics, with relevance to
varied systems ranging from cold
atoms~\cite{cold-atoms-1,cold-atoms-2} and
dissipative systems~\cite{Weiss1999} to
quantum-dot devices~\cite{Elzerman04,Petta05}
and biological donor-acceptor
molecules.~\cite{ChargeMigrationDNA2007}
Alongside fundamental question concerning
the underlying time scales and the long-time
behavior, there are many technological issues
that require careful investigation. For example,
the decoherence and relaxation of spins appears
to be the major obstacle for the realization
of quantum-computing algorithms in real
systems.~\cite{LossDiVincenzo1998} Another
key issue is the understanding of coherent
control and the switching characteristics
of nanodevices such as single-electron
transistors.~\cite{KastnerSET1992} These and
related topics require the development and
application of suitable many-body
techniques.~\cite{haugBookTransport96}

Over the years, the Kadanoff-Baym~\cite{KadanoffBaym62}
and Keldysh~\cite{Keldysh65}  techniques have
proven to be  accurate tools for describing
the real-time dynamics of weakly correlated
systems such as light-matter interaction in
semiconductors~\cite{HaugKoch2004} and the
decoherence and relaxation of an impurity
spin well above the Kondo
temperature.~\cite{LangrethWilkins1972}
Geared toward perturbation theory, these
techniques generally fail upon the development
of strong correlations, when nonperturbative
approaches are in order. A case in point are
quantum dots tuned to the Kondo
regime,\cite{NatureGoldhaberGordon1998} where
traditional diagrammatic-based approximations
are unsuitable  to describe the nonequilibrium
state.\cite{SchmittAnders2010}
The difficulty lies in the fact that strongly
correlated systems change their nature as a
function of certain control parameters such as
the temperature or the coupling constants, an
aspect well captured by renormalization-group
approaches.\cite{Wilson75,*BullaCostiPruschke2008,
Schoeller2009a} The precise status of a voltage bias as yet another
control parameter in interacting nanostructures is
still under debate.

Recent years have witnessed an impressive advancement
of numerical techniques aimed at tracking the real-time
dynamics of strongly correlated systems, primarily with
the development of the time-dependent density-matrix
renormalization group (TD-DMRG).~\cite{Marston2002,
Luo_etal2003,DaleyKollathSchollwoeckVidal2004,
*GobertTdDMRG2004,*Schollwoeck-2005,WhiteEeiguin2004}
Yet while the adaptive TD-DMRG works remarkably well
on time scales of order the reciprocal bandwidth,
it is presently unsuited for tackling longer time
scales due to an accumulated error that grows first
linearly and then exponentially with the time elapsed.
Alternative formulations~\cite{Schmitteckert2004,
*Schmitteckert2010-b} of the TD-DMRG circumvent the
accumulated error, but are simply too demanding to
advance to long times. Recent adaptations of
continuous-time Monte Carlo techniques to real-time
dynamics~\cite{Muehlbacher-rabani-2008,Werner-2009,
*Werner-2010,Weiss-et-al-2008,Schiro-Fabrizio-2009}
are free of finite-size effects, but are confined
to short time scales due to an inherent sign
problem. The Chebyshev expansion
technique,~\cite{Fehske-RMP2006} developed by Tal Ezer
and Kosloff,~\cite{TalEzer-Kosloff-84,Kosloff-94}
offers yet another extremely powerful approach for
tracking the time evolution of finite-size systems. 
However, it too is quite limited in accessing long
time scales in the presence of interactions due to
the exceedingly large Hilbert space that must be
retained. A complementary approach is provided by
the time-dependent numerical renormalization group
(TD-NRG).~\cite{AndersSchiller2005,AndersSchiller2006,
AndersSSnrg2008,*AndersNeqGf2008} The TD-NRG
can successfully bridge over vastly different
time scales, but is far more restrictive in the
systems and perturbations to which it can be applied.
A composite approach that combines the complementary
traits of the different techniques mentioned above
is highly desirable.

In this paper, we devise such a hybrid approach that
combines  the outstanding capabilities of the numerical
renormalization group (NRG) to systematically
construct the effective low-energy Hamiltonian of
the system~\cite{Wilson75,*BullaCostiPruschke2008,
Hewson2001,*HewsonEffectivParam2006} with the prowess
of complementary approaches for calculating the
real-time dynamics derived from this Hamiltonian.
Typically strongly correlated systems possess
multiple energy scales that markedly differ  in magnitude,
hence their dynamics is governed by vastly different
time scales. This spread of time scales, which may
differ by many orders of magnitude, poses an enormous
obstacle for most computational approaches. The TD-NRG
is quite unique in this respect as it can efficiently
bridge between the different time scales. Our hybrid
approach presented below provides a flexible platform for combining
the TD-NRG with one's method of choice for treating
the effective low-energy Hamiltonian. Possible
choices for the complementary method could be exact
diagonalization, the TD-DMRG, and possibly also the
time-dependent noncrossing 
approximation.~\cite{Langreth-Nordlander-1991,
ShaoLangreth1994} 
Here we shall demonstrate the applicability of our
approach by combining the TD-NRG with the Chebyshev
expansion technique (CET), providing thereby an
important proof of principle.

The basic philosophy underlying the hybrid-NRG is
to first exhaust the TD-NRG in order to decompose
the time-dependent wave function into distinct
components, each associated with a separate time
scale and evolving according to its own reduced
Hamiltonian acting on a suitable subspace
of the full Fock space.  Each of the individual
components with its associated Hamiltonian can
then be treated with improved accuracy using,
e.g., the TD-DMRG or CET. In this manner one can
exploit the successive reduction in energy scales
in order to boost the TD-DMRG and CET to long time
scales that otherwise would be inaccessible to
either of these methods. Concomitantly, the accuracy
and flexibility of the TD-NRG are greatly enhanced,
as we demonstrate by extending the approach to the
physically relevant case of repeated switchings.
The hybrid platform further offers an appealing
way to reduce discretization errors inherent to
the Wilson chain by converting to hybrid chains.

The idea to use the NRG level flow to construct
effective Hamiltonians is, of course, not
new. Dating back to the original work of
Wilson,~\cite{Wilson75,*BullaCostiPruschke2008}
this framework has been significantly advanced
by Hewson~\cite{Hewson2001} who used it as a
starting point for devising a renormalized
perturbation theory. Hewson's approach requires,
however, analytical knowledge of the form of the
low-energy Hamiltonian and its associated quantum
field theory. So far it has been applied mainly to the single-impurity
Anderson model, where it
was used, among other things, to calculate the
steady-state current~\cite{hewsonNeqSIAM05,
HewsonEffectivParam2006} in the limit of a small
bias voltage.

Our approach is far more general 
as it makes no assumption on the analytical form of
the effective Hamiltonian. Rather, it is solemnly based on
Wilson's original
concept~\cite{Wilson75,*BullaCostiPruschke2008} that
the NRG level flow contains all accessible information,
and in particular can accurately describe the crossover
region between two distinct fixed points (a regime
which generally lies outside the reach of perturbative
methods). Our framework exclusively uses the sequence
of diagonalized NRG Hamiltonians,~\cite{Wilson75,
*BullaCostiPruschke2008} circumventing thereby any
prejudice on the form of the effective Hamiltonian.
As a result our method is model independent, relying
solely on the NRG approach itself.

In this paper, we extend the original TD-NRG algorithm
from a single quantum quench to multiple switchings,
which requires an additional approximation beyond the
one underlying the conventional TD-NRG. 
As we demonstrate by explicit calculations, the quality
of the approximation can be systematically improved by
enlarging the subspace treated using the complementary
method.

\subsection{Preliminaries}
\label{sec:preliminaries}

In the TD-NRG, the continuous bath is represented
by a discretized Wilson chain,~\cite{Wilson75,
*BullaCostiPruschke2008} characterized by tight-binding
hopping matrix elements that decay exponentially
along the chain. This separation of energy scales
enables access to exponentially long time
scales~\cite{AndersSchiller2005,AndersSchiller2006}
that otherwise would be inaccessible using ordinary
tight-binding chains. In analyzing the accuracy of
the TD-NRG it is important to distinguish between
two sources of error: one extrinsic due to the
discretized representation of the continuous bath
in terms of a Wilson chain, and the other intrinsic
due to the TD-NRG algorithm for tracking the
real-time dynamics on the Wilson chain.

\begin{figure}[tbp]
\begin{center}
\hspace{-0.4cm}
\includegraphics[width=0.385\textwidth]{fig1a.eps}
\vspace{-0.7cm}

\includegraphics[width=0.46\textwidth]{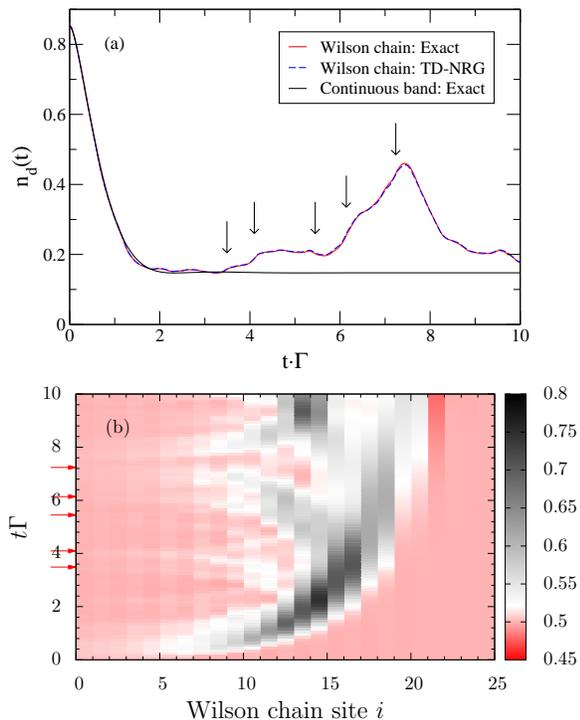}
\caption{(Color online)
         (a) Real-time dynamics of the impurity charge
         $n_d(t)$ in the resonant-level model following
         a sudden quench of the level energy from
         $E_d = -2\Gamma_0$ to $E_d = 2\Gamma_0$ at
         time $t = 0$. The red solid line shows the
         exact time evolution on the Wilson chain,
         obtained by exact diagonalization of the
         single-particle eigenmodes. The dashed blue
         line depicts the TD-NRG result, while the
         solid black line shows the exact analytical
         solution for a continuous band [given by
         Eq.~(50) of
         Ref.~\onlinecite{AndersSchiller2006}].
         (b) A two-dimensional contour plot of the
         exact time-dependent occupancies $n_i(t)$ of
         the first 26 sites along the Wilson chain,
         labeled $i = 0, \ldots, 25$. The red arrows
         indicate instances in time when reflected
         currents reach the impurity site. The very
         same times are marked by the black arrows
         in panel (a). Parameters: $\Lambda = 1.8$,
         $\Gamma_0/D = 10^{-2}$, the number of states
         kept in the TD-NRG is $N_s = 800$, and the
         chain length is $N = 40$.}
\label{fig:RLM-n-vs-t}
\end{center}
\end{figure}

As already noted in Ref.~\onlinecite{AndersSchiller2006},
the latter source of error is remarkably small. We
illustrate this point in Fig.~\ref{fig:RLM-n-vs-t}(a)
for the noninteracting resonant-level model (RLM),
describing a single fermionic level coupled by
hybridization to a conduction band (see
Sec.~\ref{sec:quench-RLM-hybrid-chain} for an
explicit definition of the model). Abruptly shifting
the energy of the level and tracking the time
evolution of the level occupancy $n_d(t)$, we
compare the TD-NRG to an exact analytical solution
for a continuous bath [given by Eq.~(50) of
Ref.~\onlinecite{AndersSchiller2006}], as well as
to an exact numerical solution on the Wilson chain
using exact diagonalization of the single-particle
eigenmodes. Only minuscule deviations are found between
the TD-NRG and the exact solution on the Wilson chain,
both of which significantly depart at some point
from the continuum-limit result. As analyzed in
Ref.~\onlinecite{AndersSchiller2006}, one can
decrease the deviations from the continuum limit
and delay them to a later time by reducing the
Wilson discretization parameter $\Lambda$. At
the same time, the deviations are hardly
affected by prolonging the chain length $N$.

This leads to two important conclusions:
(i) The main source of error in the TD-NRG
    is extrinsic rather than
    intrinsic;~\cite{AndersSchiller2006}
(ii) The exponentially decaying tight-binding
     matrix elements, which lie at the heart of the
     NRG,~\cite{Wilson75,*BullaCostiPruschke2008}
     are also the limiting factor for reproducing
     the continuum-limit result for the quench
     dynamics.

To understand the source of the deviations from
the continuum-limit result, we note that globally
conserved quantities such as the charge or the spin
of the system are locally connected by the continuity
equation to associated charge and spin currents. The
exponentially decreasing tight-binding matrix elements
along the Wilson chain significantly slow down the
propagation of such currents,~\cite{Schmitteckert2010} 
generating internal reflections at the sites of the
one-dimensional chain. This is depicted in
Fig.~\ref{fig:RLM-n-vs-t}(b), where we plot the exact
time-dependent occupancies of the first 26 sites along
the chain in response to a sudden quench of the impurity
level. The two-dimensional contour plot clearly reveals
reflections of the charge current at certain positions
and certain characteristic times. Once the reflected
currents reach the impurity site its level occupancy
starts to deviate significantly from the exact
continuum-limit result. Indeed, the red arrows in
Fig.~\ref{fig:RLM-n-vs-t}(b) indicate instances in
time when reflected charge wave fronts reach the
impurity site. At these very same times the occupancy
on the impurity level develops new structures [marked
by the black arrows in Fig.~\ref{fig:RLM-n-vs-t}(a)]
that are absent in the continuum limit. Upon
decreasing the Wilson discretization parameter
$\Lambda$ the magnitude of the reflected currents
is suppressed and the reflection points are pushed
deeper down the chain, however the effect is never
fully eliminated as long as $\Lambda > 1$.

\begin{figure}[tbp]
\begin{center}
\includegraphics[width=70mm]{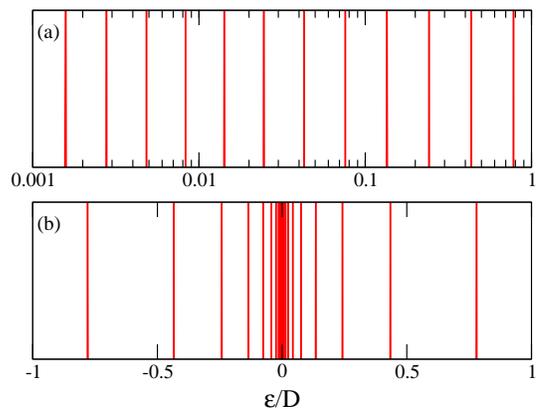}
\caption{(Color online)
         The exact single-particle energy levels
         of the RLM on a Wilson chain with
         $\Lambda = 1.8$, $\Gamma_0/D = 10^{-2}$,
         and $E_d = 0$. The chain length is $N = 40$.
         Panel (a) depicts positive energies on a
         logarithmic scale. Panel (b) shows the full
         spectrum on a linear scale.}
\label{fig:RLM-single-particle-levels}
\end{center}
\end{figure}

An alternative perspective on the effect of the Wilson
discretization procedure is provided by examining the
exact single-particle energy levels of the RLM. For
an ordinary tight-binding chain of length $N$ with
a constant hopping matrix element $\xi$, the
single-particle energy levels are roughly uniformly
distributed in the energy range $[-D, D]$, where
$D = 2\xi$ is the conduction-electron bandwidth. As
the chain length is increased the energy levels become
more densely distributed until a continuous spectrum
is recovered for $N \to \infty$. A different picture
applies to the Wilson chain. As depicted in
Fig.~\ref{fig:RLM-single-particle-levels}, the
single-particle energy levels are uniformly
distributed on a logarithmic scale, resulting in
a sparse distribution of levels at higher energies.
Enumerating the positive single-particle energy
levels from high to low, these scale as
$\epsilon_n \propto \Lambda^{-n}$, in accordance with
Wilson's logarithmic discretization of the conduction
band.~\cite{Wilson75,*BullaCostiPruschke2008} By
prolonging the chain length $N$ one increases the
total number of levels, yet the distribution of
high-energy levels retains its sparse form even
as $N \to \infty$. A continuous band is recovered
only upon implementing the combined limit
$\Lambda \to 1^+$, $N \to \infty$, which
illustrates the limitation of working with a fixed
$\Lambda > 1$.~\footnote{A related objection
has recently been raised by A.\ Rosch, who noted that
the Wilson chain cannot serve as a proper heat
reservoir even if made infinitely long; see
report no. arXiv:1110.6514.}

These argumentations clearly point to an intrinsic
tradeoff within the TD-NRG, as the very same
logarithmic discretization that enables access to
exponentially long time scales also prevents from
fully recovering the continuum limit. The question
then arises whether one can somehow reconcile these
two apparently contradicting properties, which is
one of the main goals of this work.

\subsection{Plan of the Paper}

Briefly stated, the aim of this paper is a
two-fold extension of the TD-NRG. 

The first
goal is to devise a flexible framework for
hybridizing the TD-NRG with complementary methods
of calculating real-time dynamics that do not
rely on the special structure of the Wilson
chain. By liberating ourselves from working on the
Wilson chain at all time scales, the hybrid method
should enable a systematic improvement of both
finite-size and discretization errors. 

The second
objective is to extend the method from a single
quantum quench to repeated switchings between two
distinct Hamiltonians ${\cal H}^a$ and ${\cal H}^b$.
Besides being of considerable interest on its own
right, this protocol can be viewed as a first vital
step toward treating a general time-dependent
Hamiltonian ${\cal H}(t)$. Indeed, discretizing the
time axis and replacing ${\cal H}(t)$ within each
time interval $i$ with the constant form ${\cal H}_i$,
the full time evolution can be interpreted as a
sequence of quenches. 

Despite its conceptual simplicity
this strategy has not been pursued thus far since the
initial nonequilibrium density operator $\hat{\rho}_i$
at the beginning of each time interval is neither
explicitly known nor of the analytical form required by the
original TD-NRG formulation.~\cite{AndersSchiller2006}

To this end, we derive the hybrid-NRG
in Sec.~\ref{sec:hybrid-nrg-theory}.
Since the formulation relies heavily on the TD-NRG,
we shall commence in Sec.~\ref{Sec:TD-NRG}
with a brief review of the TD-NRG following
Ref.~\onlinecite{AndersSchiller2006}. This
exposition is essential not only for keeping the
paper self-contained, but mainly for introducing the
notations and tools that will be used throughout
our construction of the hybrid-NRG. After setting
the stage in Sec.\ \ref{Sec:TD-NRG}, we present 
the first major conceptual result of this paper in 
Sec.~\ref{Sec:hybrid-TDNRG}:  the hybrid-NRG approach.
The key idea is to partition the Wilson chain into two
parts: the high-energy part is treated with the TD-NRG
while the low-energy part is feed into the complementary
method of choice. The standard TD-NRG approach can be
embedded into this more general hybrid framework by shifting
the partition to the end of the Wilson chain. In order to
make contact with the complementary approach used for
hybridizing with the TD-NRG (e.~g., the CET), we examine
the interface with the TD-NRG from a wave-function
perspective in Sec.~\ref{sec:expectation-values}.

Using the central results of Sec.~\ref{sec:hybrid-nrg-theory}, 
we extend the hybrid approach
to periodic switching between two Hamiltonians in 
Sec.~\ref{sec:Switchings}, making the simulation of coherent control
by an external field accessible to the hybrid method. This section covers 
the second major conceptual result of this paper. 
Since the exact evaluation of the reduced density matrix of
an arbitrary density operator has remained a computational
challenge, we propose a set of approximations which neglect
some of the high-energy contributions to the real-time
dynamics. These approximations are systematic and are
carefully analyzed. The complementary method used to
supplement the TD-NRG in this paper, the Chebyshev
expansion technique,\cite{TalEzer-Kosloff-84,Kosloff-94}
is reviewed in Sec.\ \ref{Sec:Chebyshev}.

Since the deviations of finite-size nonequilibrium
dynamics from the continuum limit are linked to the
bath discretization, our hybrid framework targets the
liberation of numerical simulations from the particular
form of the Wilson chain without losing access to
exponentially long time scales. In 
Sec.~\ref{sec:quench-RLM-hybrid-chain}, 
we demonstrate the potential of our hybrid framework 
by investigating the influence of different hybrid chain
types on the discretization errors encountered in the
local dynamics. In Sec.\ \ref{sec:periodic-switching},
we present our results for periodic switching. Since
the TD-NRG is embedded in our hybrid framework, we first
discuss in Sec.\ \ref{sec:td-nrg-periodic-switching}
the limitations of a simple extension of the TD-NRG
to periodic switching.
The exact solution of the finite-size RLM subject
to a periodic drive is used to benchmark both the
hybrid NRG-CET and the periodic TD-NRG.

The interacting resonant-level
model \cite{Schlottmann1980,MethaAndrei2005,BordaSchillerZawadowski2008,
BoulatSaleurSchmitteckert2008,BordaZawa2010} 
(IRLM) serves as a first nontrivial test of our hybrid
approach. It includes an additional local capacitive
coupling $U$ between the charge on the impurity level
and the fermionic band, preventing an exact solution
of its dynamics. The low-energy fixed point of the
IRLM is identical to that of the noninteracting RLM,
featuring a renormalized, $U$-dependent hybridization
strength. By comparing the real-time dynamics of both
models with identical renormalized hybridization
strengths, we show in Sec.~\ref{sec:IRLM} that the
local charge dynamics significantly deviates with
increasing $U$ from the noninteracting case.
Interaction-induced oscillations are found in the
IRLM, whose characteristic frequency depends only
on the renormalized hybridization and not directly
on $U$. Those oscillations are completely absent in
the noninteracting RLM, even though both models share
the same low-energy fixed point. Using a strong-coupling
analysis, we provide a simple physical picture for this
surprising effect.
Finally, we conclude with a discussion and outlook
in Sec.\ \ref{sec:discussion-outlook}.

\section{The hybrid-NRG}
\label{sec:hybrid-nrg-theory}

\subsection{The time-dependent NRG}
\label{Sec:TD-NRG}

The TD-NRG has been designed to track the real-time
dynamics of quantum-impurity systems following an
abrupt quantum quench. The perturbations under
consideration are implicitly assumed to be of
local character, i.e., perturbations that are
applied either to the impurity itself or to its
close vicinity. 

The Hamiltonian of a quantum-impurity system has
the generic structure
\begin{eqnarray}
{\cal H} = {\cal H}_{\rm bath} + {\cal H}_{\rm imp}
           + {\cal H}_{\rm mix} ,
\end{eqnarray}
where ${\cal H}_{\rm bath}$ models the continuous bath,
${\cal H}_{\rm imp}$ represents the decoupled impurity,
and ${\cal H}_{\rm mix}$ describes the coupling between
the two subsystems. The entire system is characterized
at time $t = 0$ by the density operator 
\begin{equation}
\hat{\rho}_0 =
     \frac{ e^{-\beta {\cal H}^i }}
          {{\rm Trace}
                \left\{
                        e^{-\beta {\cal H}^i }
                \right\} } ,
\label{rho-0}
\end{equation}
when a static perturbation $\Delta {\cal H}$ is
suddenly switched on: ${\cal H}(t \ge 0) =
{\cal H}^{i} + \Delta {\cal H} \equiv {\cal H}^{f}$.
The density operator evolves thereafter in
time according to
\begin{equation}
\hat{\rho}(t > 0) =
    e^{-i t {\cal H}^f } \hat{\rho}_0
    e^{i t {\cal H}^f } .
\label{rho-of-t}
\end{equation}
Our objective is to use the NRG to compute the
time-dependent expectation value $O(t)$ of a
general local operator $\hat{O}$. As shown in
Ref.~\onlinecite{AndersSchiller2005,AndersSchiller2006},
the result can be written in the form
\begin{eqnarray}
\langle \hat{O} \rangle (t) &=&
        \sum_{m}^{N}\sum_{r,s}^{\rm trun} \;
        e^{i t (E_{r}^m - E_{s}^m)}
        O_{r,s}^m \rho^{\rm red}_{s,r}(m) ,
\label{eqn:time-evolution-intro} 
\end{eqnarray}
where $E_{r}^m$ and $E_{s}^m$ are the dimension-full
NRG eigenenergies of the perturbed Hamiltonian at
iteration $m \le N$, $O_{r,s}^m$ is the matrix
representation of $\hat{O}$ at that iteration,
and $\rho^{\rm red}_{s,r}(m)$ is the reduced density
matrix defined in Eq.~(\ref{eqn:reduced-dm-def})
below. The restricted sum over $r$ and $s$ requires
that at least one of these states is discarded at
iteration $m$. The NRG chain length $N$ implicitly
defines the temperature entering Eq.~(\ref{rho-0}):
$T_N \propto \Lambda^{-N/2}$, where $\Lambda > 1$
is the Wilson discretization parameter.

The derivation of Eq.~(\ref{eqn:time-evolution-intro})
relies on two key ingredients:
(i) The identification of a complete basis set of
approximate NRG eigenstates for the many-body Fock
space ${\cal F}_N$ of the Wilson chain;
(ii) Expectation values are obtained by explicitly
     tracing over this complete basis set using a
     suitable resummation procedure.
Below we review these two key components
following the notations and presentation of
Ref.~\onlinecite{AndersSchiller2006}.

\begin{figure}[tbp]
\centering
\includegraphics[width=0.48\textwidth]{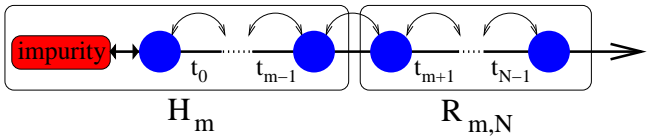}
\caption{\small
         The full Wilson chain of length $N$ is
         divided into a subchain of length $m$
         and the ``environment'' $R_{m,N}$. The
         Hamiltonian ${\cal H}_m$ can be viewed
         either as acting only on the subchain
         of length $m$, or as acting on the full
         chain of length $N$, but with the hopping
         matrix elements $t_m, \cdots, t_{N - 1}$ all
         set to zero. The former picture is the
         traditional one. In the TD-NRG one adopts
         the latter point of view.} 
\label{fig:semi-infinite-nrg}
\end{figure}

\subsubsection{Complete basis set}
\label{sec:complete-basis-set}

The NRG targets an iterative solution of a quantum
impurity coupled to a finite Wilson chain with $N$
chain links.~\footnote{Even though the chain has
$N+1$ bath sites we refer to it either as an $N$-site
chain or a chain of length $N$, so as to emphasize
the number of distinct hopping matrix elements or
energy scales involved.} Similar to the initial
sweep in the finite-size DMRG, one can view the NRG
procedure as a set of operations, where at first all
hopping matrix elements are set to zero along the
$N$-site chain, and at each successive step another
hopping matrix element is switched on. The full
Hamiltonian ${\cal H}_N$ is recovered once all
hopping matrix elements have been switched on.
The entire sequence of Hamiltonians ${\cal H}_m$
with $m \leq N$ act in this picture on the same
Fock space ${\cal F}_N$ of the $N$-site chain, hence
each NRG eigenenergy of ${\cal H}_m$ has an extra
degeneracy of $d^{(N-m)}$, where $d$ is the number
of distinct configurations at each site along the
chain. The extra degeneracy stems from the $N-m$
``environment'' sites at the end of the chain, denoted
by $R_{m, N}$ in Fig.~\ref{fig:semi-infinite-nrg},
which remain decoupled from the impurity at
iteration $m$.

When acting on the $m$-site chain, we label the NRG
eigenstates and eigenenergies of ${\cal H}_m$ by
$\{ |r; m\rangle \}$ and $E_{r}^{m}$, respectively.
Consider now the action of ${\cal H}_m$ on the full
$N$-site chain. Enumerating the different configurations
of site $i$ by $\{ \alpha_i \}_{i = 1, \ldots, d}$,
each of the tensor-product states $| r; m \rangle
\otimes | \alpha_{m + 1}, \ldots, \alpha_N \rangle$
with arbitrary $\alpha_{m+1}, \ldots, \alpha_N$ is a
degenerate NRG eigenstate of ${\cal H}_m$ with energy
$E_{r}^{m}$. To label these states we introduce the
shorthand notation $\ket{r,e;m}$, where the
``environment'' variable
$e = \{\alpha_{m+1}, \ldots,\alpha_N\}$ encodes the
$N - m$ site labels $\alpha_{m+1}, \ldots, \alpha_N$,
and the index $m$ is used to record where the chain is
partitioned into a ``subsystem'' and an ``environment''
(see Fig.~\ref{fig:semi-infinite-nrg}).

In order to retain a manageable number of states, the
high-energy eigenstates are discarded after each
iteration, which is fully justified in equilibrium
by the hierarchy of energy scales along the Wilson
chain and the Boltzmannian form the equilibrium
density operator. Regarding all states of the final
iteration as discarded, it has been show in
Refs.~\onlinecite{AndersSchiller2005,AndersSchiller2006}
that the collection of all states discarded in the
course of the NRG iterations form a complete basis
set of approximate NRG eigenstates for the full
$N$-site chain. 

To understand this important point,
consider the first iteration $m_{\rm min}$ at which
states are discarded. In order to keep track of
the complete basis set of the $N$-site chain, the
eigenstates $\ket{r,e;m_{\rm min}}$ can be formally
divided into two distinct subsets: the discarded
high-energy states
$\{ \ket{l,e;m_{\rm min}}_{\rm dis} \}$ and the kept
low-energy states $\{ \ket{k,e;m_{\rm min}}_{\rm kp} \}$. 
Obviously, the sum of the two subsets form a complete
basis set of the full chain. To simplify the notations
we shall omit hereafter the subscripts $_{\rm dis}$
and $_{\rm kp}$, and will use in exchange the indices
$l$ and $k$ to label the discarded and kept states,
respectively. At the next NRG iteration only the kept
states are used to construct the NRG eigenstates of
${\cal H}_{m_{\rm min}+1}$ within the truncated
subspace spanned by $\{ \ket{k,e;m_{\rm min}} \}$.
The resulting NRG eigenstates can again be divided
into two subsets of discarded and kept states which,
when combined with the discarded eigenstates of
iteration $m_{\rm min}$, form a complete basis set
of the Fock space ${\cal F}_N$ of the full $N$-site
chain. Repeating this procedure at all subsequent
iterations, one continues to maintain a complete
basis set of ${\cal F}_N$ up to the final NRG
iteration $m = N$.  In this manner we arrive at
the following completeness relation
\begin{eqnarray}
\sum_{m = m_{\rm min}}^N \sum_{l,e}
     \ket{l,e;m}\bra{l,e;m} &=& 
     \sum_{m = m_{\rm min}}^N 
     \hat{P}_{m} = 1 ,
\label{equ:complete-basis}
\end{eqnarray}
where the summation over $m$ starts from the first
iteration $m_{\rm min}$ at which a basis-set
reduction is imposed. Here the summation indices
$l$ and $e$ implicitly depend on $m$, and the
projector onto the subspace discarded at iteration
$m_{\rm min}\le m\le  N$ is defined as
\begin{equation}
\hat{P}_{m} = \sum_{l, e}
              | l, e; m \rangle \langle l, e; m| \; .
\label{P_m}
\end{equation}

The complementary projector $\hat{1}^{+}_{m}$ onto
the subspace retained at iteration $m$ ($m < N$)
is given in turn by
\begin{equation}
\hat{1}^{+}_{m} =
    \sum_{k, e}
         | k, e; m \rangle \langle k, e; m| \ ,
\label{1^+_m}
\end{equation}
which can be recast in the form
\begin{equation}
\hat{1}^{+}_{m} =
    \sum_{m' = m + 1}^{N} \hat{P}_{m'} \; .
\label{eqn:p-kept-states}
\end{equation}
This latter equality reflects the fact that all
states retained at iteration $m$ are necessarily
discarded at some later iteration $m'$. In
particular, since all states of the final iteration
$N$ are regarded discarded, then $\hat{1}^{+}_{N}$
is identically zero while $\hat{1}^{+}_{N - 1}$
coincides with $\hat{P}_{N}$. Combined with
Eq.~(\ref{eqn:p-kept-states}), the completeness
relation of Eq.~(\ref{equ:complete-basis})
can further be partitioned into
\begin{equation}
1 = \sum_{m = m_{\rm min}}^{M} \hat{P}_m
  + \hat{1}_{M}^{+}
\label{Completeness-2}
\end{equation}
with arbitrary $m_{\rm min} \leq M \leq N$. This
useful identity will be repeatedly used in
constructing the hybrid-NRG.

\subsubsection{Reduced density matrix and the TD-NRG
               algorithm}

Following a sudden quench, the time evolution of
the system is governed by the perturbed Hamiltonian
${\cal H}^f$ while the initial condition is encoded
in the initial density matrix $\hat{\rho}_0$. For
quantum-impurity systems, all relevant information
on the initial condition is contained in
Eq.~(\ref{eqn:time-evolution-intro}) in the form of
the reduced density matrices~\cite{AndersSchiller2005,
AndersSchiller2006} $\rho^{\rm red}_{s,r}(m)$,
defined as
\begin{equation}
\rho^{\rm red}_{s,r}(m) = \sum_{e}
          \langle s,e;m|\hat{\rho}_0 |r,e;m \rangle .
\label{eqn:reduced-dm-def}
\end{equation}
Here the states $|r,e;m \rangle$ and $|s,e;m \rangle$
correspond to the Hamiltonian ${\cal H}^f$, and the
summation runs over the environment degrees of
freedom $e$. The only approximation entering 
Eq.~(\ref{eqn:time-evolution-intro}) is the standard
NRG approximation ${\cal H}_N |r,e;m \rangle
\approx {\cal H}_m |r,e;m \rangle = E_r^m |r,e;m \rangle$,
which enables us to write
\begin{equation}
\langle s,e;m|\hat{\rho}(t) |r,e;m \rangle
     = e^{i t (E_{r}^m - E_{s}^m)}
       \langle s,e;m|\hat{\rho}_0 |r,e;m \rangle .
\end{equation}
Apart from this sole point,
Eq.~(\ref{eqn:time-evolution-intro})
constitutes an exact evaluation of $O(t)$ on
the discretized $N$-site chain. 

Practical calculations hinge on the ability to
accurately compute the reduced density matrices
of Eq.~(\ref{eqn:reduced-dm-def}). For a general
$\hat{\rho}_0$ this can be a daunting task. However,
in the case of interest where $\hat{\rho}_0$ has
the standard Boltzmann form of Eq.~(\ref{rho-0}),
the summation over $e$ can be carried out exactly.
Hence $\rho^{\rm red}_{s,r}(m)$ can be evaluated
at the same level of accuracy as the equilibrium
density operator $\hat{\rho}_0$. Technically this
goal is achieved by implementing two independent
NRG runs, one for the initial Hamiltonian
${\cal H}^i$ in order to construct $\hat{\rho}_0$
using to Eq.~(\ref{rho-0}), and another for
the full Hamiltonian ${\cal H}^f$. The reduced
density matrix $\rho^{\rm red}_{s,r}(m)$ is first
evaluated with respect to the eigenstates of the
initial Hamiltonian, and then
rotated~\cite{AndersSchiller2005,AndersSchiller2006}
to the eigenstates of the full Hamiltonian using
the overlap matrices
\begin{eqnarray}
\langle q_i; m | r; m \rangle =
           S_{q_i,r}(m) .
\label{eq:S_kr-def}
\end{eqnarray}
Here $| r; m \rangle$ denotes an NRG eigenstate
of the full Hamiltonian at iteration $m$, and
$| q_i; m \rangle$ is an NRG eigenstate of the
initial Hamiltonian at the same iteration. The
method of calculating the overlap matrices
$S_{q_i,r}(m)$ is detailed in
Ref.~\onlinecite{AndersSchiller2006}.

\subsection{Derivation of the Hybrid-NRG}
\label{Sec:hybrid-TDNRG}

The original TD-NRG approach, summarized above,
tracks the quench dynamics of a quantum-impurity
system in terms of the phase factors
$e^{i t (E_{r}^m - E_{s}^m)}$ and the reduced
density matrices $\rho^{\rm red}_{s,r}(m)$
assigned to each NRG iteration $m$. 
Although quite elegant and useful,
it is less transparent how to incorporate
ideas from methods such as the TD-DMRG or CET,
as these deal with wave functions directly.
To develop a convenient and flexible interface
between the TD-NRG and these vastly different
approaches we  reformulate the
former approach from a wave-function perspective.

\subsubsection{Wave-function formulation}

Let us commence with accurately stating the
problem from a wave-function perspective. We are
interested in tracking the time evolution of some
initial state $|\psi_0 \rangle$ under the dynamics
defined by the Hamiltonian ${\cal H}$ acting on a
finite Wilson chain of length $N$. We shall not
concern ourselves at this stage with how the
initial state $|\psi_0 \rangle$ is generated, but
will elaborate on this important point later on.

Formally our task boils down to computing
\begin{equation}
| \psi(t) \rangle =
      e^{-i {\cal H} t} | \psi_0 \rangle \; .
\end{equation}
Application of the completeness relation of
Eq.~(\ref{Completeness-2}) to the state
$| \psi(t) \rangle$ leads to its partitioning
according to
\begin{equation}
| \psi(t) \rangle =
      \sum_{m = m_{\rm min}}^{M} \!
           | \phi_m (t) \rangle
      + | \chi_M (t) \rangle \; ,
\label{partitioning-of-psi}
\end{equation}
where
\begin{equation}
| \phi_m (t) \rangle = \hat{P}_m | \psi(t) \rangle
\label{phi_m_t}
\end{equation}
and
\begin{equation}
|\chi_M (t) \rangle = \hat{1}_{M}^{+} |\psi(t) \rangle
\label{chi_m_t}
\end{equation}
are the projections of $| \psi(t) \rangle$
onto the subspaces defined by $\hat{P}_m$
and $\hat{1}_{M}^{+}$, respectively.
Equation~(\ref{partitioning-of-psi}) simply
converts the general state $| \psi(t) \rangle$
into a concrete representation in terms of 
our complete basis set.

\subsubsection{Evaluation of expectation values}
\label{sec:expectation-values}

Given the time-evolved wave function of 
Eq.~(\ref{partitioning-of-psi}), we proceed to compute
time-dependent averages of physical observables:
\begin{eqnarray}
A(t) & =& \langle \psi(t) | \hat{A}|\psi(t) \rangle
\; .
\label{eqn:def-expection-A}
\end{eqnarray}
To this end, we use the completeness relation of
Eq.~(\ref{Completeness-2}) to decompose any arbitrary
operator $\hat{A}$ into
\begin{eqnarray}
\hat{A} &=&
        \sum_{m, m'}^{M}
             \hat{P}_m \hat{A} \hat{P}_{m'}
        + \sum_{m}^{M}
               \left \{
                        \hat{P}_m \hat{A} \hat{1}^{+}_{M}
                        + \hat{1}^{+}_{M} \hat{A} \hat{P}_m
               \right \}
\nonumber \\
        &+& \hat{1}^{+}_{M} \hat{A} \hat{1}^{+}_{M} \; ,
\label{A-via-Pm}
\end{eqnarray}
where the summations over $m$ and $m'$ start from
$m_{\rm min}$. Writing the first two terms on the
right-hand side of Eq.~(\ref{A-via-Pm}) as
\begin{eqnarray}
&& \sum_{m = m_{\rm min}}
        \left [
                \hat{P}_m \hat{A} \hat{P}_{m}
                + \hat{P}_m \hat{A}
                \left (
                        \sum_{m' = m+1}^{M}
                             \hat{P}_{m'}
                        + \hat{1}^{+}_{M}
                \right )
        \right .
\nonumber \\
&&
   \;\;\;\;\;\;\;\;\;\;\;\;\;\;\;
        + \left .
                  \left (
                          \sum_{m' = m+1}^{M}
                               \hat{P}_{m'}
                          + \hat{1}^{+}_{M}
                  \right )
                  \hat{A} \hat{P}_m
          \right ]
\end{eqnarray}
and noting that
\begin{equation}
\hat{1}^{+}_m = \sum_{m' = m+1}^{M} \hat{P}_{m'}
        + \hat{1}^{+}_{M} \; ,
\end{equation}
the operator $\hat{A}$ is recast in the exact form
\begin{equation}
\hat{A} = \sum_{m = m_{\rm min}}^{M} \hat A(m)
        + \hat A_\chi \; ,
\label{eqn:operator-projection}	
\end{equation}
where
\begin{equation}
\hat A(m) = \hat{P}_{m} \hat A \hat{P}_{m}
            + \hat{1}_{m}^{+} \hat A \hat{P}_{m}
            + \hat{P}_{m} \hat A \hat{1}_{m}^{+}
\end{equation}
and
\begin{equation}
\hat{A}_\chi = \hat{1}_{M}^{+} \hat A \hat{1}_{M}^{+} \; .
\end{equation}
Here the index $M$ can take any value in the range
$m_{\rm min} \le M \le N$. Explicitly, the operator
$\hat{A}(m)$ has the formal representation
\begin{equation}
\hat A(m) =
     \sum_{r,s}^{\rm trun}
     \sum_{e,e'} 
          | r, e; m \rangle
          \langle r, e; m | \hat{A} |s, e'; m \rangle
          \langle s, e'; m | \; ,
\label{A-of-t-2}
\end{equation}
where the restricted sum $\sum_{r, s}^{\rm trun}$
implies, as before, that at least one of the states
$r$ and $s$ is discarded at iteration $m$.

As in the original TD-NRG, we focus hereafter on local
operators $\hat{A}$ that act solely on degrees of
freedom that reside either on the impurity itself
or on the first $m_{\rm min}$ sites along the Wilson
chain.~\cite{AndersSchiller2006} For any such local
operator, the matrix elements in Eq.~(\ref{A-of-t-2}) are
diagonal in and independent of the environment variables
$e$ and $e'$:
\begin{equation}
\langle r, e; m | \hat{A} |s, e'; m \rangle
        = A^{m}_{r, s} \delta_{e,e'} \; . 
\end{equation}
Substituting the operator decomposition of
Eq.~(\ref{eqn:operator-projection}) into
Eq.~(\ref{eqn:def-expection-A}) and using the definition
$| \chi_M(t) \rangle = \hat{1}_{m}^{+} |\psi(t) \rangle$
of Eq.~(\ref{chi_m_t}), the time-dependent expectation
value takes the form
\begin{equation}
A(t) = \sum_{m = m_{\rm min}}^{M}
            \langle \psi(t)|\hat{A}(m)|\psi(t) \rangle
     + \langle \chi_M(t) | \hat{A} | \chi_M(t) \rangle
\; .
\end{equation}
This general expression reduces for a local operator to
\begin{equation}
A(t) = \langle \chi_M(t) | \hat{A} | \chi_M(t) \rangle
      + \! \sum_{m = m_{\rm min}}^{M} \!
           \sum_{r, s}^{\rm trun}
                       A^{m}_{r, s} \rho_{s, r}^{m}(t)
\; ,
\label{A-of-t-3}
\end{equation}
where
\begin{equation}
\rho_{s, r}^{m}(t) =
     \sum_e
         \langle s, e; m | \psi(t) \rangle
         \langle \psi(t)| r, e; m \rangle
\label{rho-m-of-t}
\end{equation}
is the reduced density matrix at iteration $m$.

Three comments should be made about Eqs.~(\ref{A-of-t-3})
and (\ref{rho-m-of-t}). First, these expressions are
both general and exact for the real-time dynamics on
the discretized chain. Apart from the restriction to
local operators, no further approximations or
assumptions are involved. Second, Eqs.~(\ref{A-of-t-3})
and (\ref{rho-m-of-t}) can be easily extended to a
statistical admixture of initial states
$\{ |\psi_i \rangle \}$ with the statistical weights
$\{ w_i \}$. This requires the simple substitutions
\begin{equation}
\langle \chi_M(t) | \hat{A} | \chi_M(t) \rangle \to
        \sum_i w_i
                   \langle \chi_{M, i}(t) | \hat{A}
                           | \chi_{M, i}(t) \rangle
\end{equation}
and
\begin{equation}
\hat{\rho}(t)= | \psi(t) \rangle \langle \psi(t)| \to 
    \sum_{i}
             w_i |\psi_{i}(t)\rangle \langle\psi_{i}(t)|
\end{equation}
in Eqs.~(\ref{A-of-t-3}) and (\ref{rho-m-of-t}),
respectively. Third, the conventional TD-NRG approach
is recovered from Eqs.~(\ref{A-of-t-3}) and
(\ref{rho-m-of-t}) by 
(i) setting $M = N$,
(ii) realizing that $\hat{A}_\chi = 0$ for $N = M$, and
(iii) adopting the standard NRG approximation
${\cal H}|r, e; m\rangle  \approx E_r^m |r, e; m\rangle$,
which simplifies $\rho_{s, r}^{m}(t)$ to
$e^{i ( E^m_r - E^m_s) t} \rho^{\rm red}_{s, r}(m)$ with
\begin{equation}
\rho^{\rm red}_{s, r}(m) =
     \sum_e
         \langle s, e; m | \psi_0\rangle
         \langle \psi_0| r, e; m \rangle \; .
\end{equation}

A natural generalization of the TD-NRG is to apply the
NRG approximation ${\cal H}|r, e; m\rangle \approx
E_r^m |r, e; m\rangle$ to the early iterations
$m \leq M$ only, converting Eqs.~(\ref{A-of-t-3}) and
(\ref{rho-m-of-t}) to
\begin{eqnarray}
A(t) &=& \sum_{m = m_{\rm min}}^{M} \!
         \sum_{r, s}^{\rm trun}
                     e^{i ( E^m_r - E^m_s) t}
                     A^{m}_{r, s}
                     \rho_{s, r}^{\rm red}(m)
\non
  && + \langle \chi_M(t) | \hat{A} | \chi_M(t) \rangle
\; .
\label{hybrid}
\end{eqnarray}
This equation, which constitutes one of the central
results of this paper, interpolates between the
TD-NRG, corresponding to $M = N$, and the exact
time-dependent expectation value, obtained for
$M = m_{\rm min}$. Of course, the latter statement
assumes an exact evaluation of $| \chi_M(t) \rangle$,
which is an impractical task for $M = m_{\rm min}$.
As discussed below, a proper choice of the parameter
$M$ allows for an improved evaluation of
$| \chi_M(t) \rangle$ using alternative methods such 
as the TD-DMRG or CET, with minimal loss of accuracy
at the early iterations to which the NRG approximation
is applied. Furthermore, by resorting to methods
that do not rely on the special structure of the
Wilson chain to evaluate $| \chi_M(t) \rangle$, one
can abandon the exponential decay of the hopping
matrix elements beyond site $M$, reducing thereby
the discretization errors inherent to the Wilson
chain. These principles form the core of the hybrid
approach. We now turn to elaborate on the technicalities
of how $M$ is selected, the interface with the
hybridized method, and the way in which the initial
state $| \psi_0 \rangle$ is constructed.

\subsection{Interface between the TD-NRG and the
            hybridized approach}

\subsubsection{Hierarchy of energy scales and the time
               evolution of $| \chi_M(t) \rangle$}

To turn Eq.~(\ref{hybrid}) into an operative platform
for hybridizing the TD-NRG with alternative methods
of computing the real-time dynamics of
$| \chi_M(t) \rangle$, it is useful to go back to
the partitioning of $| \psi(t) \rangle$ specified in
Eq.~(\ref{partitioning-of-psi}) and gain a deeper
insight into the energy scales encoded into the
projectors $\hat{P}_m$. Applying the operator
decomposition of Eq.~(\ref{eqn:operator-projection})
to the Hamiltonian ${\cal H}$, the latter is written as
\begin{eqnarray}
{\cal H} &=& \sum_{m = m_{\rm min}}^{M} \H(m)
           + \hat{h}_\chi \; ,
\label{eqn:H-operator-projection}
\end{eqnarray}
where
\begin{equation}
{\cal H}(m) \equiv  \hat{P}_m {\cal H} \hat{P}_{m}
              + \hat{P}_m {\cal H} \hat{1}^{+}_m
              + \hat{1}^{+}_m {\cal H} \hat{P}_{m}
\end{equation}
and
\begin{equation}
\hat{h}_\chi = \hat{1}^{+}_M {\cal H} \hat{1}^{+}_M \; .
\end{equation}
It is rather easy to see that the different
Hamiltonian terms that appear in
Eq.~(\ref{eqn:H-operator-projection})
generally do not commute with one another,
i.e., $[\H(m), \hat{h}_{\chi} ] \neq 0$
and $[\H(m), \H(m')]\neq 0$ if $m\neq m'$.
According to the NRG philosophy, however, the
off-diagonal terms $\hat{P}_{m} {\cal H} \hat{P}_{m'}$
with $m \neq m'$ are expected to be small, as
these couple excitations on different energy
scales. Consequently, one can approximate
$\H(m)$ with $\hat h_m = \hat{P}_{m} {\cal H}
\hat{P}_{m}$ to obtain the approximate Hamiltonian 
\begin{equation}
{\cal H} \approx \sum_{m = m_{\rm min}}^{M} \hat h_m
           + \hat h_\chi \; .
\label{NRG-approx-1}
\end{equation}

Evidently, Eq.~(\ref{NRG-approx-1}) becomes exceedingly
more accurate the smaller is $M$, acquiring the
status of an identity for $M = m_{\rm min}$,
since ${\cal H} =  \hat h_\chi$ in this case.
Furthermore, since the Hamiltonian terms $\hat{h}_\chi$ and
$\hat{h}_m$ with $m \leq M$ are confined to
the subspaces projected out by $\hat{1}^{+}_M$
and $\hat P_m$, respectively, the Hamiltonian of
Eq.~(\ref{NRG-approx-1}) is block-diagonal in these
subspaces with $[\hat{h}_m , \hat{h}_{m'}] =
[\hat{h}_m , \hat{h}_{\chi}] = 0$. This allows us
to write the time-dependent state $| \psi(t) \rangle$
within this approximation as
\begin{equation}
| \psi(t) \rangle =
          \sum_{m = m_{\rm min}}^{M} \!
                e^{-i \hat  h_m t}| \phi_m \rangle
          + e^{-i \hat  h_{\chi} t}| \chi_M \rangle \; ,
\label{decomposition}
\end{equation}
where $| \phi_m \rangle = \hat{P}_m | \psi_0 \rangle$
and $| \chi_M \rangle = \hat{1}_M^{+} | \psi_0 \rangle$
are the projections of the initial state onto
the subspaces defined by $\hat{P}_m$ and
$\hat{1}_{M}^{+}$, respectively. In other terms, it
suffices in this approximation to first project out 
$| \chi_M \rangle$ and $| \phi_m \rangle$ from the
initial state, and then propagate them separately
in time, each according to its own Hamiltonian.  

Physically, Eq.~(\ref{decomposition}) prescribes a
decomposition of the desired time-dependent state
into independent components, each associated with
a different time scale $t_m = 1/D_m \sim \Lambda^{m/2}$
and evolving according to its own reduced Hamiltonian
(either $\hat h_m$ or $\hat h_{\chi}$). In the case
of spinless electrons the reduced Hamiltonian
$\hat h_m$ has the explicit form
\begin{eqnarray}
\hat h_m &=& \sum_{l, e}
              E_l^m |l, e; m \rangle \langle l, e; m |
\non
              &&
    + \sum_{n = m}^{N - 1}
              t_n \{
                     \hat{P}_m
                     f_{n + 1}^{\dagger} f_n
                     \hat{P}_m
                     + {\rm H.c.}
                  \} \; ,
\label{h_m-explicit}
\end{eqnarray}
where $f_{n}^{\dagger}$ creates an electron on
the $n$th site of the Wilson chain, $t_n$ is the
dimension-full hopping matrix element between
sites $n$ and $n+1$ along the chain, $l$ runs over
the NRG eigenstates discarded at iteration $m$, and
$E_l^m$ denotes their corresponding NRG eigenenergies.
Note that the projection operators $\hat{P}_m$
in the right-most term are attached in practice
only to $f_m$ and $f^{\dagger}_m$, as all other
operators $f_n$ with $n > m$ do not possess any
matrix element that takes us out of the subspace
defined by $\hat{P}_m$. The Hamiltonian
$\hat h_{\chi}$ is nearly identical, except that
the index $m$ is replaced with $M$ and the discarded
states $|l, e; m \rangle$ are replaced with the
NRG eigenstates retained at the conclusion of
iteration $M$. In the presence of additional bands
the Wilson orbitals $f_{n}^{\dagger}$ acquire an
additional flavor index $\nu$, which may label the
spin $\sigma$, an orbital channel $\alpha$, or the
spin-channel tuple $\nu=(\sigma,\alpha)$ as in
two-channel Kondo models (see, e.g.,
Ref.~\onlinecite{LebanonSchillerAnders2003}). Other
than setting $f_{n}^{\dagger}\to f_{n\nu}^{\dagger}$
and adding a suitable summation over $\nu$, the very
same equations carry over to the general multiband
case.

\subsubsection{Physical role of the parameter $M$}

The Hamiltonian of Eq.~(\ref{h_m-explicit}) can be
interpreted as modeling a hyper-impurity with the
localized configurations $|l \rangle$ and eigenenergies
$E_l^{m}$, which are tunnel-coupled to a chain of length
$N - m$. The size of the impurity is equal to the number
of states discarded at iteration $m$. Thus, the calculation
of $|\phi_m (t) \rangle = e^{-i h_m t}| \phi_m \rangle$
becomes exceedingly more affordable the larger is $m$
due to the exponential reduction of the Fock space of
the chain $R_{m, N}$ attached to the hyper-impurity. 
We stress, however, that the dimension of the
subspace associated with $\hat{P}_{m_{\rm min}}$
is comparable in size to that of the full Wilson
chain, hence an accurate evaluation of
$|\phi_{m_{\rm min}} (t) \rangle$ is similar in
complexity to the calculation of the full state
$| \psi (t) \rangle$.

There is little computational gain in implementing
Eq.~(\ref{decomposition}) if all components of
the wave function must be accurately computed. 
Fortunately, this generally is not the case for the
class of problems of interest, where $| \psi_0 \rangle$
is some low-lying  eigenstate (typically the ground
state) of an initial Hamiltonian ${\cal H}^{i}$. Under
these circumstances $| \psi_0 \rangle$ typically has
only negligible overlap with the high-energy states of
${\cal H}$, i.e., $\langle \phi_m |\phi_m \rangle \ll 1$
for the initial NRG iterations. Overlap becomes
significant only upon approaching a characteristic
energy scale $D_M$ where the spectra of the full and
the unperturbed Hamiltonians begin to notably deviate
from one another. Usually this happens at some
characteristic low-energy scale of the problem,
e.g., the Kondo temperature $T_K$ in case of the
Kondo Hamiltonian.
 
Consequently, the initial NRG iterations with $m \leq M$
can be treated using further approximations such as
setting $t_n = 0$ in Eq.~(\ref{h_m-explicit}),
corresponding to the standard NRG approximation.
Since the reduced density matrix $\rho_{s, r}^{m}(t)$
requires only the matrix element
$ \langle \psi(t)| r, e; m \rangle$, one can
implement $e^{i\H t} | r, e; m \rangle$ instead of
propagating $| \phi_m(t) \rangle$ in time, which
simplifies the exact result of Eq.~(\ref{A-of-t-3}) to
the approximate expression of Eq.~(\ref{hybrid}). The
computational effort can therefore be focused on
evaluating $| \chi_M(t) \rangle$, which dominates
the expectation value $A(t)$. Most importantly, given
the initial state $| \chi_M \rangle$ and the effective
Hamiltonian $\hat{h}_{\chi}$ generated by the NRG,
$| \chi_M(t) \rangle$ can be computed using one's
method of choice.

\begin{figure}[btp]
\centering
\includegraphics[width=0.48\textwidth]{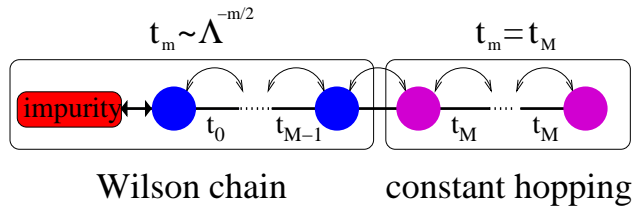}
\caption{\small
         A hybrid Wilson chain where the first
         $M + 1$ hopping matrix elements decrease
         according to $t_n \propto \Lambda^{-n/2}$
         with $\Lambda > 1$, and all further
         hopping matrix elements are held constant
         and equal to $t_M$.}
\label{fig:nrg-tight-binding}
\end{figure}

The physical role of the integer $M$, which so far
served as a mere parameter, is now disclosed:
it defines the NRG iteration $M$ beyond which the
time-dependent state should be accurately computed.
Moreover, this partitioning can be used to improve
the discrete representation of the continuous bath
as noted above. For example, one can design a
hybrid chain such that all sites up to $m = M$ are
discretized with $\Lambda > 1$ and all further
sites are converted to $\Lambda \to 1^+$ (see
Fig.~\ref{fig:nrg-tight-binding}). Such a chain is
impractical for pure NRG-based calculations, but is
made possible by resorting to alternative methods
for tracking the time evolution of
$| \chi_M(t) \rangle$. In this manner
discretization errors are significantly reduced
at the energy scale $D_M$, corresponding to the
time scale $t_M = 1/D_M$. The crucial point to notice 
is that the reduced
Hamiltonian $h_{\chi}$ has the effective bandwidth
$D_M\propto \Lambda^{-M/2} \ll D$ and acts on a
reduced chain of length $N - M$. This enables access
to long time scales of order $t_M \gg 1/D$ using
techniques such as the TD-DMRG or CET, which otherwise
are restricted to far shorter times.

The only remaining uncertainty pertains to a suitable
choice of the iteration number $M$. In the absence of a
sharp mathematical criterion, the choice of $M$ should
be considered on a case-by-case basis. Qualitatively,
one expects the scale $D_M$ to correspond to
$\max \{ \Gamma_0, |\epsilon_d| \}$ for the resonant-level
model [see Eq.~(\ref{H-NI-RLM}) below], and to the
Kondo temperature~\cite{Wilson75,*BullaCostiPruschke2008}
$T_K$ for the Kondo model. The case of an Anderson
impurity is clearly more subtle, as spin and charge
relax on different time scales.~\cite{AndersSchiller2005}
Here a different optimal choice of $M$ may 
apply to observables acting on the spin and charge
sectors.

\subsubsection{Construction of the initial state
               $\ket{\psi_0}$}
\label{sec:initial-psi-0}

So far, we have assumed the decomposition of the
initial state $\ket{\psi_0}$ according to
Eq.~(\ref{partitioning-of-psi}), but did not
specify how $| \psi_0 \rangle$ is obtained in
practice. The construction of $|\psi_0\rangle$
and its projection $| \chi_M \rangle$ onto the
low-energy subspace defined by $\hat{1}^{+}_M$
depends in detail on the method hybridized with
the TD-NRG. Our discussion below covers both the
TD-DMRG and CET.

We begin with the initial NRG run, which
provides us with the low-energy Hamiltonian
$\hat h^{i}_{\chi} =\hat{1}^{+}_{i, M} {\cal H}^i
\hat{1}^{+}_{i, M}$ corresponding to the initial
Hamiltonian ${\cal H}^{i}$. Here $\hat{1}^{+}_{i, M}$
denotes the projection operator onto the low-energy
subspace of ${\cal H}^{i}$ retained at the
conclusion of iteration $M$. As detailed in
Eq.~(\ref{h_m-explicit}), $\hat h^{i}_{\chi}$
comprises of a hyper-impurity, a residual chain of
length $N - M$, and a tunnel coupling between both
parts of the system. In the next step the ground
state $| \psi_0 \rangle$ of $\hat h^{i}_{\chi}$
is computed. In case of the TD-DMRG this is done
using the standard DMRG
algorithm,~\cite{White-92,* White-93} while for
the CET (for which a shorter chain $R_{M, N}$ is
employed) the Davidson method~\cite{Davidson} can be
used. At the conclusion of this step one has the initial
state $|\psi_0 \rangle$ at hand, expressed via the
kept NRG eigenstates of ${\cal H}^{i}_{M}$:
\begin{equation}
| \psi_0 \rangle =
         \sum_{k_i, e}
              c_{k_i, e} |k_i, e; M \rangle \; .
\end{equation}

Given $| \psi_0 \rangle$, the state
$| \chi_M \rangle$ is obtained by projecting
$| \psi_0 \rangle$ onto the low-energy subspace
of the full Hamiltonian defined by $\hat{1}^{+}_M$.
This in turn yields
\begin{equation}
| \chi_M \rangle =
         \sum_{k, e}
              b_{k, e} |k, e; M \rangle
\end{equation}
with
\begin{equation}
b_{k, e} = \sum_{k_i}
                S^{\ast}_{k_i, k}(M) c_{k_i, e} \; ,
\end{equation}
where $S(M)$ is the overlap matrix defined in
Eq.~(\ref{eq:S_kr-def}). This state is then
propagated in time according to $\ket{\chi_M(t)} =
e^{-i \hat h_{\chi}t } |\chi_M \rangle $ using
either the TD-DMRG or CET and fed into
Eq.~(\ref{hybrid}). As for the reduced density
matrices $\rho_{s, r}^{\rm red}(m)$ entering
Eq.~(\ref{hybrid}), these are computed recursively
from $| \psi_0 \rangle$ using the standard TD-NRG
algorithm.~\cite{AndersSchiller2006}

\section{Repeated switchings}
\label{sec:Switchings}

Armed with the hybrid-NRG platform, we proceed
in this section to the second major result of this
paper --- the extension of the TD-NRG from a single
quantum quench to repeated switching events. 

Coherent
control of small nanodevices such as semiconductor
quantum dots or superconducting flux qubits can be
achieved by applying gate-voltage pulses or
time-dependent electromagnetic fields. For example,
circularly polarized laser pulses are used to
induce and control the spin polarization in
semiconductor quantum dots.~\cite{GreilichBayer2007}
Alternatively, one can apply rapid changes to
close-by gate voltages in order to control the energy
levels and/or the tunneling rates of a quantum dot.
Each of these protocols involves repeated switchings
between two distinct configurations of the applied
fields, which we denote hereafter by $a$ and $b$. 
Theoretically this corresponds to periodic conversions
in time between two quantum-impurity Hamiltonians,
${\cal H}^a$ and ${\cal H}^b$, that differ in those
components describing the isolated dot and its coupling
to the leads. Our goal is to track the time evolution
of local expectation values in response to such a
sequence of switching events.

To clearly formulate the problem, we assume that
the system resides at time $t = 0$ in a low-lying
eigenstate of the Hamiltonian ${\cal H}^{a}$, when
its Hamiltonian is abruptly converted from
${\cal H}^{a}$ to ${\cal H}^b$. The system then
evolves in time under the influence of ${\cal H}^b$
up to time $\tau > 0$, when the Hamiltonian of the
system is suddenly switched back to ${\cal H}^a$
for the duration $\tau < t < 2\tau$. This sequence of
switchings is repeated periodically with a period of
$2 \tau$, as described by the time-dependent Hamiltonian
\begin{equation}
{\cal H}(t > 0) =
         \left \{
                  \begin{array}{ll}
                       {\cal H}^{a}\;\;\;\;\;\; &
                       (2n - 1) \tau \le t < 2n \tau
                       \\
                       {\cal H}^{b} &
                       2n \tau \le t <  (2n + 1) \tau
                  \end{array}
         \right .
\end{equation}
(here $n = 0, 1, 2,\ldots$). Within each time interval
where the Hamiltonian is constant, the expectation
value of a general local operator $\hat{A}$ is given
by Eqs.~(\ref{A-of-t-3}) and (\ref{rho-m-of-t})
where, depending on the time interval in question,
$|\chi_M(t)\rangle$ and $\rho^{m}_{s,r}(t)$ pertain
either to ${\cal H}^a$ or ${\cal H}^b$. Explicitly,
$|\chi_M(t)\rangle$ is replaced with
$|\chi_M^{(\alpha)} (t)\rangle =
1^{+}_{\alpha, M} |\psi(t) \rangle$,
where $1^{+}_{\alpha, M}$ denotes the projection
operator of Eq.~(\ref{1^+_m}) written with respect
to NRG eigenstates of ${\cal H}^{\alpha}$
($\alpha = {\rm a}, {\rm b}$). With these adjustments
the resulting expression is formally exact, but requires
explicit knowledge of the states $|\psi(t) \rangle$
and $|\chi_M^{(\alpha)}(t) \rangle$.

Consider a particular time interval
$(2n - 1) \tau < t < 2n \tau$ in which the system
evolves according to the Hamiltonian ${\cal H}^a$.
As discussed in Sec.~\ref{sec:expectation-values},
the expectation value $A(t)$ can be approximated by
applying the conventional NRG approximation to the
early iterations $m \leq M$ only. This yields
Eq.~(\ref{hybrid}), where $|\chi_M(t)\rangle$, the
eigenenergies $E_r^m$ and $E_s^m$, and the state
labels $s$ and $r$ correspond to ${\cal H}^a$. The
only formal modification as compared to
Eq.~(\ref{hybrid}) pertains to the reduced density
matrix $\rho^{\rm red}_{s, r}(m)$, in which the
initial state $|\psi_0\rangle$ must be replaced with 
$|\psi (t) \rangle$ at time $t_{2n -1} = (2n - 1) \tau$.
The same set of rules carry over to any given time
interval $2n \tau < t < (2n + 1) \tau$, except that
${\cal H}^a$ is replaced with ${\cal H}^b$, and the
initial state $|\psi_0\rangle$ is replaced with
$|\psi (t) \rangle$ at time $t_{2n} = 2n \tau$. This
leaves us with those instances in time where the
Hamiltonian is abruptly converted from ${\cal H}^a$
to ${\cal H}^b$ or vice versa.

For concreteness let us focus on the time instance
$t_{2n} = 2n \tau$, when the Hamiltonian is converted
from ${\cal H}^a$ to ${\cal H}^b$. As described in
Eq.~(\ref{partitioning-of-psi}), the state of the
system can be formally decomposed within the time
interval $(2n - 1) \tau < t < 2n \tau$ into
\begin{equation}
|\psi(t) \rangle =
         |\chi_M^{\rm (a)}(t) \rangle
         + |\delta \psi^{\rm (a)}(t) \rangle
\end{equation}
with $|\delta \psi^{\rm (a)}(t) \rangle =
\sum_{m \leq M} |\phi_m^{\rm (a)}(t) \rangle$. Here the
states $|\phi_m^{\rm (a)}(t) \rangle$ are projected
according to the NRG eigenstates of ${\cal H}^a$. The
time-dependent density operator is next divided into
$\hat{\rho}_{\chi}^{\rm (a)}(t) +
\delta \hat{\rho}^{\rm (a)}(t)$, where
\begin{equation}
\hat{\rho}_{\chi}^{\rm (a)}(t) =
           |\chi_M^{\rm (a)}(t) \rangle
           \langle \chi_M^{\rm (a)}(t)|
\end{equation}
and $\delta \hat{\rho}^{\rm (a)}(t) = \hat{\rho}(t) -
\hat{\rho}_{\chi}^{\rm (a)}(t)$. Setting $t \to t_{2n}$ and
replacing $\hat\rho_0 = |\psi_0 \rangle \langle \psi_0 |
\to |\psi(t_{2n}) \rangle \langle \psi(t_{2n}) | $
in Eq.~(\ref{eqn:reduced-dm-def}), the reduced density
matrix with respect to the eigenstates of ${\cal H}^{a}$
reads
\begin{eqnarray}
\rho^{\rm red \, (a)}_{\, s, r}(m) &=&
     \underbrace
     {
          \sum_{e}
               \langle s,e;m|
                       \hat{\rho}_\chi^{\rm (a)}(t_{2n})
               |r,e;m \rangle
     }_{\chi_{s, r}^{\rm (a)}(m)}
\non &&
     + \delta \rho_{s, r}^{\rm (a)}(m) ,
\label{rho-red-via-chi}
\end{eqnarray}
where we have defined
\begin{eqnarray}
\delta \rho_{s, r}^{\rm (a)}(m) &\equiv &
          \sum_{e}
               \langle s, e;m|
                       \delta \hat{\rho}^{\rm (a)}(t_{2n})
               |r, e;m \rangle
\; .
\end{eqnarray}
An equivalent expression
$\rho^{\rm red \, (b)}_{\, s, r}(m) =
\chi^{\rm (b)}_{\, s, r}(m) +
\delta \rho^{\rm (b)}_{\, s, r}(m)$ applies to the
reduced density matrix with respect to the NRG
eigenstates of the Hamiltonian ${\cal H}^b$.

In the notations of Ref.~\onlinecite{AndersSchiller2006}
[see Eqs.(31) and (32) therein], the first term
$\chi^{\rm (a)}(m)$ in Eq.~(\ref{rho-red-via-chi})
is of the pure form $\chi^{++}(m)$, having none of
the components $\chi^{+-}(m)$, $\chi^{-+}(m)$, and
$\chi^{--}(m)$. Thus, one has the exact
conversion~\cite{AndersSchiller2006}
\begin{equation}
\chi^{\rm (b)}(m) = S^{\dagger}(m) \,
                    \chi^{\rm (a)}(m) \, S(m)  \; ,
\label{trans-of-chi}
\end{equation}
where $S(m)$ is the overlap matrix between the
NRG eigenstates of ${\cal H}^a$ and ${\cal H}^b$
defined by Eq.~(\ref{eq:S_kr-def}). By contrast,
the second term $\delta \rho^{\rm (a)}(m)$ in
Eq.~(\ref{rho-red-via-chi}) involves all four
components $\delta \rho^{++}(m)$, $\delta \rho^{+-}(m)$,
$\delta \rho^{-+}(m)$, and $\delta \rho^{--}(m)$,
and as a result lacks an explicit
relation~\cite{AndersSchiller2006} to
$\delta \rho^{\rm (b)}(m)$. Similar to
Eq.~(\ref{trans-of-chi}), we approximate
$\delta \rho^{\rm (b)}(m)$ by
\begin{equation}
\delta \rho^{\rm (b)}(m) = S^{\dagger}(m) \,
             \delta \rho^{\rm (a)}(m) \, S(m) \; ,
\label{delta-rho-transformed}
\end{equation}
leading to the compact transformation rule
\begin{equation}
\rho^{\rm red \, (b)}(m) = S^{\dagger}(m) \,
               \rho^{\rm red\, (a)}(m) \, S(m) \; .
\label{rho-a-to-rho-b}
\end{equation}
The complementary transformation rule for switching
from ${\cal H}^b$ to ${\cal H}^a$ reads
\begin{equation}
\rho^{\rm red \, (a)}(m) = S(m) \,
      \rho^{\rm red\, (b)}(m) \, S^{\dagger}(m) \; .
\label{rho-b-to-rho-a}
\end{equation}

It should be emphasized that
Eq.~(\ref{delta-rho-transformed}) is an {\em ad-hoc}
approximation, whose quality is difficult to assess
{\em a-priori}. Its  accuracy  is necessarily controlled
by the smallness of $\langle \delta \psi^{(\alpha)} |
\delta \psi^{(\alpha)} \rangle$, which in turn is
reduced the smaller is $M$. Below we present numerical
results demonstrating this point.

In addition to the reduced density matrix,
Eq.~(\ref{hybrid}) requires explicit knowledge of the
projected state, either $|\chi_M^{\rm (a)}(t) \rangle$
or $|\chi_M^{\rm (b)}(t) \rangle$, depending on the
time interval in question. It is therefore necessary to
formulate the transformation rule of $|\chi_M(t) \rangle$
upon conversion of the Hamiltonian. Focusing again
on the time instance $t = t_{2n}$, the state
$|\chi_M^{\rm (b)}(t_{2n}) \rangle$ is formally given by
\begin{widetext}
\begin{equation}
|\chi_M^{\rm (b)}(t_{2n}) \rangle
    = \hat{1}^{+}_{b, M} |\psi(t_{2n}) \rangle
    = \sum_{m = m_{\rm min}}^{M} \!
             \hat{1}^{+}_{b, M}
             | \phi_m^{\rm (a)} (t_{2n}) \rangle
             + \hat{1}^{+}_{b, M}
               | \chi_M^{\rm (a)}(t_{2n}) \rangle \; ,
\label{chi_2n-switch}
\end{equation}
\end{widetext}
where $| \phi_m^{\rm (a)} (t_{2n}) \rangle$ and
$| \chi_M^{\rm (a)}(t_{2n}) \rangle$ are projected
according to the operators $\hat{P}_{\, a, m}$ and
$\hat{1}^{+}_{a, M}$ pertaining to the Hamiltonian
${\cal H}^a$. The right-most term in
Eq.~(\ref{chi_2n-switch}) has the exact representation
\begin{equation}
\hat{1}^{+}_{b, M}
        | \chi_M^{\rm (a)} (t_{2n}) \rangle
    = S^{\dagger}(M)
        | \chi_M^{\rm (a)} (t_{2n}) \rangle \; ,
\end{equation}
which follows from the fact that
$\hat{P}_{\, a, m} | \chi_M^{\rm (a)}(t_{2n}) \rangle$
vanishes by construction for all $m \le M$. As
for the remaining terms on the right-hand side of
Eq.~(\ref{chi_2n-switch}), these require explicit
knowledge of the states
$| \phi_m^{\rm (a)} (t_{2n}) \rangle$.
Unfortunately, it is unfeasible to keep track of the
states $| \phi_m^{\rm (a)}(t) \rangle$, which forces yet
another approximation. Our strategy is to omit these
terms altogether, with the expectation that their combined
contribution is typically small. Naturally, the quality
of this approximation will depend on the choice of $M$
and on details of ${\cal H}^a$ and ${\cal H}^b$. This
leaves us with the approximate transformation rule
\begin{equation}
| \chi_M^{\rm (b)} (t_{2n}) \rangle
       \approx
               S^{\dagger}(M)
               | \chi_M^{\rm (a)}(t_{2n}) \rangle \; ,
\end{equation}
along with its counterpart
\begin{equation}
| \chi_M^{\rm (a)} (t_{2n + 1}) \rangle
       \approx S(M)
               | \chi_M^{\rm (b)} (t_{2n+1}) \rangle \; .
\end{equation}

We are now in position to summarize the proposed
algorithm for computing
$A(t) = \langle \psi(t) | \hat{A} | \psi(t) \rangle$
for a sequence of switching events. We begin by
rewriting Eq.~(\ref{hybrid}) in the form
\begin{eqnarray}
A(t) &=& \sum_{m = m_{\rm min}}^{M} \!
            \sum_{r, s}^{\rm trun}
                 A^{m, (\alpha)}_{r, s}
                 \rho_{s, r}^{{\rm red}\, (\alpha)}(m; t)
\non
&&     + \langle \chi_M^{(\alpha)}(t) | \hat{A}
                 | \chi_M^{(\alpha)}(t) \rangle \; ,
\label{A(t)-switching}
\end{eqnarray}
where $\alpha$ equals ${\rm a}$ or ${\rm b}$
depending on the time interval, and
$A^{m, (\alpha)}_{r, s}$ is the matrix representation
of $\hat{A}$ at iteration $m$ with respect to the NRG
eigenstates of ${\cal H}^{\alpha}$. The first step is
to evaluate $\rho_{s, r}^{\rm red \, (a)}(m; t = 0)$
and $| \chi_M^{\rm (a)} (t = 0) \rangle$ using the
TD-NRG methodology detailed in
Ref.~\onlinecite{AndersSchiller2006}. From this point
on these quantities are evolved in time according to
the following set of rules:
\\
(i) At time $t=t_{2n}$ we switch from $\H^a$ to $\H^b$
    by setting
\begin{eqnarray}
    \rho^{\rm red \, (b)}(m; t_{2n}) &=&
            S^{\dagger}(m)\,
            \rho^{\rm red\,(a)}(m; t_{2n}) \, S(m) \; ,
\non
    | \chi_M^{\rm (b)} (t_{2n}) \rangle &=&
               S^{\dagger}(M)
               | \chi_M^{\rm (a)} (t_{2n}) \rangle \; .
\label{chi_M-switching-1}
\end{eqnarray}
(ii) In the time interval $t_{2n} \leq t \leq t_{2n + 1}$,
the density matrices and wave function are propagated
in time according to
\begin{eqnarray}
    \rho^{\rm red \, (b)}_{s, r}(m; t) &=&
       \rho^{\rm red \, (b)}_{s, r}(m; t_{2n}) \,
       e^{i ( E^m_{s} - E^m_{r}) (t - t_{2n})} \; ,
\non
    | \chi_M (t) \rangle &=&
       e^{-i\hat h_\chi^{b} (t - t_{2n})}
       \chi_M^{\rm (b)} (t_{2n}) \rangle \; ,
\end{eqnarray}
where $\hat h_\chi^{b}$ is the projected low-energy
Hamiltonian corresponding to ${\cal H}^b$.
\\
(iii) At time $t=t_{2n+1}$ we switch back from
$\H^b$ to $\H^a$ by setting
\begin{eqnarray}
    \rho^{\rm red \, (a)}(m; t_{2n+1}) &=&
        S(m)\, \rho^{\rm red\,(b)}(m; t_{2n+1}) \,
        S^\dagger (m) \; ,
\non
    | \chi_M^{\rm (a)} (t_{2n+1}) \rangle &=&
        S(M) | \chi_M^{\rm (b)} (t_{2n+1}) \rangle \; .
\label{chi_M-switching-2}
\end{eqnarray}
(iv) The density matrices and wave function are
propagated in the time interval $t_{2n + 1} \leq t
\leq t_{2n + 2}$ according to
\begin{eqnarray}
    \rho^{\rm red \, (a)}_{s, r}(m; t) &=&
        \rho^{\rm red \, (a)}_{s, r}(m; t_{2n+1}) \,
        e^{i ( E^m_{s} - E^m_{r}) (t - t_{2n+1})} \; ,
\non
    | \chi_M (t) \rangle &=&
        e^{-i\hat h_\chi^{a} (t - t_{2n+1})}
       \chi_M^{\rm (a)} (t_{2n+1}) \rangle \; ,
\label{eqn:chi-rho-evolution-2}
\end{eqnarray}
where $\hat h_\chi^{a}$ is the projected low-energy
Hamiltonian corresponding to ${\cal H}^a$.

We stress that four distinct approximations enter
Eqs.~(\ref{chi_M-switching-1})--(\ref{eqn:chi-rho-evolution-2}):
(i) A discrete representation of the continuous bath;
(ii) The standard NRG approximation
     ${\cal H}_N |r,e;m \rangle \approx
     E_r^m |r,e;m \rangle$ that is used to propagate
     the reduced density matrices within each time
     interval where ${\cal H}$ is fixed;
(iii) The transformation rule of
      Eq.~(\ref{delta-rho-transformed}) for
      $\delta \rho(m)$ at each switching event; and
(iv) The omission of those terms that originate from
     the $|\phi_m\rangle$'s in Eq.~(\ref{chi_2n-switch})
     in the transformation rule for $|\chi_M(t)\rangle$
     at each switching event.
While the first two approximations are general in
nature and apply, in particular, to the TD-NRG, the
latter two approximations are specific to the present
formulation of repeated switchings. The quality of those
approximations, which become exact for $M = m_{\rm min}$,
can be tested {\em a-posteriori} by comparison to
cases where exact solutions are available, as
done in Sec.~\ref{sec:RLM-switching} for the
noninteracting resonant-level model.

\section{Chebyshev expansion technique}
\label{Sec:Chebyshev}

Our presentation thus far was quite general and did not
specify a particular method to be hybridized with the
TD-NRG. As a proof of principle, we shall demonstrate
in Sec.~\ref{sec:periodic-switching}  the hybridization
of the TD-NRG with the CET, whose principles and
implementation are reviewed below. 

The CET~\cite{TalEzer-Kosloff-84,Kosloff-94,Fehske-RMP2006}
offers an accurate way to calculate the time evolution
of an initial state $|\psi_0 \rangle$ under the
influence of a general stationary finite-dimensional
Hamiltonian ${\cal H}$:
\begin{align}
|\psi(t) \rangle = e^{-i{\cal H}t}|\psi_0\rangle .
\label{stationarySolution}
\end{align}
The main idea of the method is to construct a stable
numerical approximation for the time-evolution
operator $e^{-i{\cal H}t}$ that is independent of
the initial state $|\psi_0\rangle$ and whose
error can be reduced to machine precision. Its
limitation lies in the need to explicitly store
certain states in the course of the calculation,
which limits the size of the Hilbert space that
can be handled. 

There are different ways to expand the time-evolution
operator, the most direct one being the conventional
expansion of the exponent in powers of ${\cal H}$.
One would like, however, to use an expansion that
converges uniformly, independent of the initial
state $|\psi_0\rangle$. A suitable choice are the
Chebyshev polynomials, defined by the recursion
relation
\begin{equation}
T_{n+1}(z) = 2zT_n(z)-T_{n-1}(z) ,
\end{equation}
subject to the initial conditions $T_0(z) = 1$ and
$T_1(z) = z$. As is well known, the Chebyshev polynomials
can be used to expand any function $f(z)$ on the interval
$-1 \leq z \leq 1$. Explicitly, $f(z)$ is expressed as
an infinite series
\begin{align}
f(z) = \sum_{n=0}^{\infty} b_n T_n(z) ,
\label{f-z-expantion}
\end{align}
where the expansion coefficients $b_n$ are given by
\begin{align}
b_n = \frac{2 - \delta_{n, 0}}{\pi}
      \int_{-1}^{1}
              \frac{f(x)T_n(x)}{\sqrt{1-x^2}} dx .
\label{b_n_integral}
\end{align}
The expansion in terms of Chebyshev polynomials can be
equally applied to any function $F(z)$ with an arbitrary
support $\lambda_{\rm min} \le z \le \lambda_{\rm max}$
using the transformation
\begin{align}
z' = 2\frac{z- \lambda_{\rm min}}
           {\lambda_{\rm max}-\lambda_{\rm min}} - 1 ,
\end{align}
which maps the interval $\lambda_{\rm min} \le z \le
\lambda_{\rm max}$ onto $-1 \leq z' \leq 1$. In doing
so one expands in practice the function $f(z') = F(z)$.

Using the rules laid above, the function $e^{-iz}$ is
expanded for $\lambda_{\rm min} \le z
\le \lambda_{\rm max}$ as
\begin{equation}
e^{-iz} = \sum_{n = 0}^{\infty} b_n T_n(z')
\end{equation}
with
\begin{subequations}
\begin{eqnarray}
b_0 &=& e^{-i\varphi}
      J_0 \left (
                  \frac{\Delta \lambda }{2}
          \right ) ,
          \\
b_{n > 0} &=& 2 i^n e^{-i\varphi}
            J_n \left (
                        \frac{\Delta \lambda }{2}
                \right ) .
\end{eqnarray}
\end{subequations}
Here $J_{n}(x)$ are the Bessel functions, $\varphi$
equals $(\lambda_{\rm max} + \lambda_{\rm min})/2$, and
$\Delta \lambda = \lambda_{\rm max} - \lambda_{\rm min}$.
Accordingly, the time-evolution operator
$e^{-i {\cal H} t}$ is expanded as
\begin{equation}
e^{-i {\cal H}t} =
      \sum_{n = 0}^{\infty} b_n(t) T_n({\cal H}') ,
\label{Chebyshev-exp-e^-iH}
\end{equation}
where
\begin{subequations}
\label{b_n}
\begin{eqnarray}
b_0(t) &=& e^{-i\alpha t}
      J_0 \left (
                  \frac{\Delta E t }{2}
          \right ) ,
\\
b_{n > 0}(t) &=& 2 i^n e^{-i\alpha t}
            J_n \left (
                        \frac{\Delta E t}{2}
                \right ) .
\end{eqnarray}
\end{subequations}
Here $E_{\rm max}$ ($E_{\rm min}$) is the maximal
(minimal) eigenenergy of ${\cal H}$,
$\Delta E$ equals $E_{\rm max} - E_{\rm min}$,
$\alpha = (E_{\rm max} + E_{\rm min})/2$, and
${\cal H}'$ is the ``transformed'' Hamiltonian
\begin{align}
{\cal H}' = 2\frac{{\cal H}- E_{\rm min}}
                  {E_{\rm max}-E_{\rm min}} - 1 .
\end{align}
Finally, applying Eq.~(\ref{Chebyshev-exp-e^-iH})
to the initial state $|\psi_0\rangle$ one obtains
\begin{equation}
|\psi(t) \rangle = \sum_{n = 0}^{\infty}
                          b_n(t) |\phi_n \rangle ,
\label{CET}
\end{equation}
where the infinite set of states
$|\phi_n \rangle = T_n({\cal H}') |\psi_0\rangle$
obey the recursion relation\cite{Fehske-RMP2006}
\begin{equation}
|\phi_{n+1} \rangle = 2{\cal H}' |\phi_n \rangle
                    - |\phi_{n-1} \rangle ,
\label{chebyshev-recursion-relation}
\end{equation}
subject to the initial condition
$|\phi_0 \rangle = |\psi_0 \rangle$ and
$|\phi_1 \rangle = {\cal H}' |\psi_0 \rangle$.

Several comments are in order. First, all time
dependence is confined in Eq.~(\ref{CET}) to
the expansion coefficients $b_n(t)$ of
Eqs.~(\ref{b_n}), which are independent of the
initial state $|\psi_0\rangle$.
Second, the Chebyshev recursion relation of
Eq.~(\ref{chebyshev-recursion-relation}) reveals
the iterative nature of the calculations. Starting
form the initial state $|\psi_0\rangle$, one constructs
all subsequent states $|\phi_n\rangle$ using repeated
applications of the ``transformed'' Hamiltonian
${\cal H}'$. Note that in practice only two such
states need be stored in memory at each given time.
Third, since $J_n(x) \sim (e x/2 n)^n$ for large
order $n$, the Chebyshev expansion converges
quickly as $n$ exceeds $\Delta E t$. Finally, the
Chebyshev expansion has the virtue that numerical
errors are practically independent of $t$, allowing
access to extremely long times. The main limitation
of the approach, as commented above, stems from the
size of the Hilbert space, since each of the states
$|\phi_n\rangle$ must be constructed explicitly.
In our applications of the approach (where
$| \chi_M \rangle$ serves as the initial state), this
Hilbert space comprises of the kept NRG states at
iteration $M$ --  typically of the order of $2^{10}$ -- 
and the remaining chain $R_{M, N}$, whose dimension
is $d^{N-M}$ ($d = 2$ being the number of distinct
configurations of a single spinless site). As a result,
application of the CET is confined to rather short
chains that cannot be used to access arbitrarily
long time scales.

\section{Single-quench dynamics on a hybrid chain}
\label{sec:quench-RLM-hybrid-chain}

As already pointed out in the introduction, the
Wilsonian discretization of the continuous bath
significantly influences the quench dynamics,
independent of which finite-size approach --- exact
diagonalization, TD-NRG, or the TD-DMRG --- is used
to track the real-time dynamics of the system. As
illustrated in Fig.~\ref{fig:RLM-n-vs-t}, deviations
from the exact continuum-limit result stem from current
reflections at sites along the Wilson chain, caused by
the exponentially decreasing hopping matrix elements
that are used. These in turn produce an exponentially
decreasing transport velocity.

In this section, we present a preliminary discussion
aimed at demonstrating the potential of substituting
the standard Wilson chain with a hybrid chain of the
type depicted in Fig.~\ref{fig:nrg-tight-binding}.
We shall do so by investigating a single quantum
quench in the noninteracting resonant-level model
(RLM), which can be solved exactly on essentially any
finite-size chain using exact diagonalization of the
single-particle eigenmodes. The availability of an
exact analytical solution for the real-time dynamics
of the level occupancy in the continuum limit [see
Eq.~(50) of Ref.~\onlinecite{AndersSchiller2006}] makes
this model an ideal benchmark for testing the quality
of different hybrid chains in reproducing the
continuum-limit result. Thus, one can clearly separate
deviations caused by the structure of the chain from
those that originate from further approximations
underlying the TD-DMRG or TD-NRG. Furthermore, the
RLM sets the stage for more complicated interacting
models, such as the interacting resonant-level model
discussed below.

Physically, the RLM describes the coupling of a single
spinless level to a continuous band of width $2D$ via
the single-particle hopping matrix element $V$:
\begin{equation}
{\cal H} = \sum_{k} \epsilon_k c^{\dagger}_{k} c^{}_{k}
         + E_d d^{\dagger} d
         + \frac{V}{\sqrt{N_k}} \sum_{k}
                  \bigl \{
                           c^{\dagger}_{k} d
                           + {\rm H.c.}
                  \bigr \} .
\label{H-NI-RLM}
\end{equation}
Here $d^{\dagger}$ creates an electron on the level,
$c^{\dagger}_{k}$ creates a band electron with
momentum $k$, and $N_k$ labels the number of distinct
values of $k$. The relevant energy scales in the
problem include the level energy $E_d$, along with
the hybridization width $\Gamma_0 = \pi \varrho V^2$,
where $\varrho$ is the conduction-electron density
of states at the Fermi level.

\begin{figure}[tb]
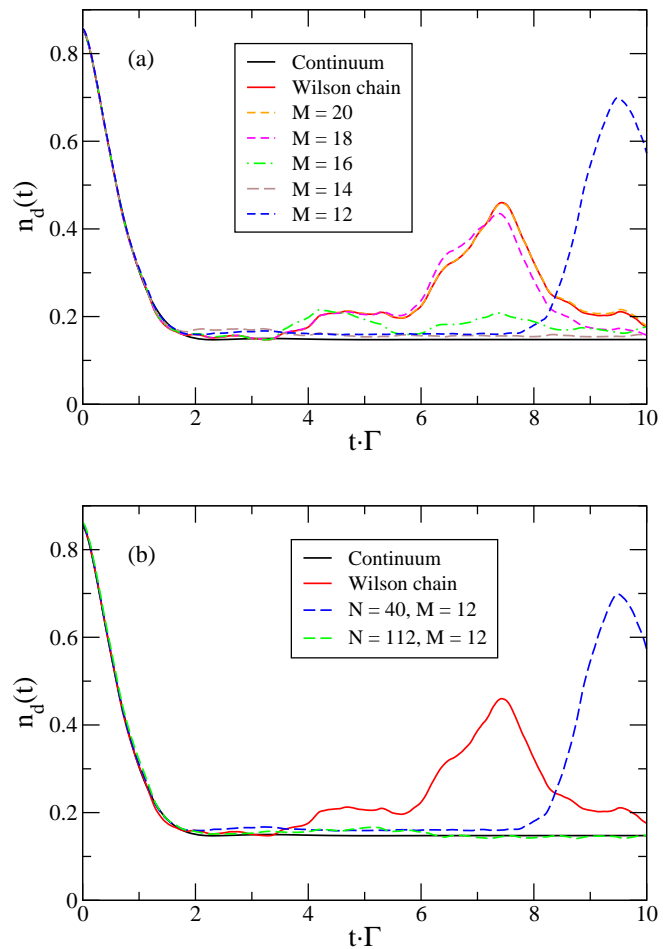

\centering 

\includegraphics[width=0.48\textwidth]{fig5a.eps}
\vspace*{2.5mm}

\includegraphics[width=0.48\textwidth]{fig5b.eps}
\vspace*{-1mm}

\caption{(Color online) Comparison of the exact
         continuum-limit solution for $n_d(t)$ in the
         RLM (solid black line) and its exact numerical
         evaluation for various hybrid chains of the type
         depicted in Fig.~\ref{fig:nrg-tight-binding}.
         The full red line depicts the case of a pure
         Wilson chain of length $N = 40$.
	 Panel (a): Keeping the total chain length
         fixed at $N = 40$ and varying the partitioning
         parameter $M$.
	 Panel (b): For $M = 12$ and two different
         chain lengths $N = 40$ and $112$. Model
         parameters: $\Gamma_i = \Gamma_f = \Gamma_0$,
         $E_d^{f} = -E_d^{i} = 2\Gamma_0$,
         $\Gamma_0/D = 10^{-2}$, and $\Lambda=1.8$.
         }
\label{fig:RLM-hybrid-chain-quench}
\end{figure}

We are interested in tracking the real-time
dynamics of the level occupancy
$n_d(t) = \langle d^{\dagger}(t) d(t) \rangle$ 
in response to a sudden quench from 
$(E_d^{i}, \Gamma_i)$ to  $(E_d^{f}, \Gamma_f)$.
Figure~\ref{fig:RLM-hybrid-chain-quench} depicts
the time evolution of $n_d(t)$ after the level
energy has been abruptly shifted from
$E_d^{i} = -2\Gamma_0$ to $E_d^{f} =2\Gamma_0$
while keeping the hybridization width fixed at
$\Gamma_0$. We used the same model parameters as
in Fig.~\ref{fig:RLM-n-vs-t}, but varied the chain
structure by considering different values of the
partitioning parameter $M$ and different chain
lengths $N$. Explicitly, as illustrated in
Fig.~\ref{fig:nrg-tight-binding}, these chains
comprise of an initial Wilson chain with
exponentially decreasing hopping matrix elements
$t_m\propto \Lambda^{-m/2}$ up to $m = M$, converting
to a constant hopping matrix element $t_m = t_M$
for $M \le m \le N$.

As can be seen in
Fig.~\ref{fig:RLM-hybrid-chain-quench}(a), there is a
systematic improvement in the agreement with the
continuum-limit result upon decreasing $M$. Specifically,
for $M = 12$ the deviations remain quite small up to
$t\cdot \Gamma \sim 8$, at which point there is a revival
of $n_d(t)$ which nearly reaches its original value
$n_d(t = 0)$ for $t \cdot \Gamma \approx 9.5$. In contrast
to a pure Wilson chain, the current velocity becomes
a constant along the chain sites with $m > M$, hence
current reflections occur only at the end of the
chain.~\cite{Schmitteckert2004}
The time at which the
revival of $n_d(t)$ is observed is simply given by the
round-trip time for a charge pulse that originates from
the impurity to reach the chain end and return.
Note that this time is shorter for $M = 12$ than
for $M = 14$.

Since charge is a globally conserved quantity, true
thermalization and relaxation can occur only in the
thermodynamic limit $N \to \infty$.  The deviation of the 
total charge from its thermalized
value  following the quench is simply given by the difference
in the equilibrated charges before and after the quench.
Since the change in total charge is ${\cal O}(1)$ for
such a local quench, there is an ${\cal O} (1/N)$
contribution to each reservoir degree of freedom  
which can be safely 
neglected in the thermodynamic limit. 

For any finite-size chain, however, this effect
remains finite and relevant for the long-time limit.
In particular, the continuity equation leads to charge
reflections at the end of the tight-binding
chain~\cite{Schmitteckert2004,Schmitteckert2010} such
that the round-trip time is controlled by the chain length.
Indeed, in Fig.~\ref{fig:RLM-hybrid-chain-quench}(b)
we compare  $n_d(t)$ for two hybrid chains, each
partitioned at $M = 12$. The two chains share the
same characteristic energy scale $D_M$ for their
partitioning, but one has a total length of $N = 40$
and the other $N = 112$. While the short-time dynamics is
nearly identical and agrees well with the continuum-limit
result, the revival time for the long hybrid chain
is pushed way beyond the time interval presented in
Fig.~\ref{fig:RLM-hybrid-chain-quench}. Thus,
discretization errors have been nearly eliminated on
the time scale of interest.

This latter example vividly illustrates the potential
of hybridizing the TD-NRG with the TD-DMRG, as the
tight-binding chain length involved is of typical DMRG
size. Since in our example the effective bandwidth $D_M$
is of order $\Gamma_0$, the TD-NRG generates an
effective Hamiltonian $h_\chi$ whose effective bandwidth
$D_M$ can be of several orders of magnitude smaller
than the bare conduction-electron bandwidth $D$. This
in turn allows access to long time scales of order
$1/\Gamma_0 \gg 1/D$ which lie beyond the reach of
pure TD-DMRG, while significantly reducing discretization
errors that are inherent to the TD-NRG.

A concluding word is in order about the optimal choices
of $M$ and the structure of the hybrid chain. The answers
to these questions are quite difficult and can be expected
to depend both on the model and observable of interest.
For instance, spin and charge excitations generally
propagate with different  velocities. It is therefore quite
feasible that the optimal choices of $M$ and $N$  will
dependent on the observable in question. A more
systematic study of the optimal hybrid chain is left
for future research.

\section{Periodic switching}
\label{sec:periodic-switching}

\subsection{Periodic switching in the
            resonant-level model}
\label{sec:RLM-switching}

So far, we have stressed the potential of using
hybrid chains by comparing their exact numerical
solutions with the continuum-limit result for
simple quenches. In this section, we extend our
focus to the hybrid-NRG approach as formulated in
Sec.~\ref{sec:Switchings} for periodic switching. To
benchmark our switching algorithm, we compare it with
exact diagonalization solutions on a pure Wilson chain,
postponing further discussion of the usage of hybrid
chains and the comparison to exact continuum-limit
results. Therefore, all calculations are performed
on a standard Wilson chain with identical parameters,
hybridizing the CET with the TD-NRG at a given site
$M$. In order to illustrate how the effective
low-energy Hamiltonian generated by the NRG can be
fed into the CET (or into any other method of choice
for that matter), we consider an extreme wide-band
limit by setting the bare bandwidth of the RLM to
$D = 10^{5}\Gamma_0$.

Using the notations introduced in
Sec.~\ref{sec:Switchings}, we start at time
$t = 0$ with a system that resides in the ground
state of ${\cal H}^a$, and switch repeatedly between
${\cal H}^a$ and ${\cal H}^b$ after each additional
time interval $\tau$. As our basic energy scale we set
$\Gamma_0 = 10^{-5} D$, whose associated time scale
$1/\Gamma_0 = 10^5/D$ lies many orders of magnitude
beyond the reach of the CET when applied directly to
the full Wilson chain. We work with a chain of fixed
length $N = 84$ and the logarithmic discretization
parameter $\Lambda = 1.4$, such that
$D_N \sim \Gamma_0/10$.

To initiate the calculation, we first perform an NRG
run for the initial Hamiltonian ${\cal H}^a$ and
select the ground state of the $N$th iteration as our
initial state~\footnote{For a hybrid chain, one
can no longer use the NRG to generate the ground
state of ${\cal H}^a$. In this case one can use the
Davidson method~\cite{Davidson} to construct the initial
state, as discussed in Sec.~\ref{sec:initial-psi-0}.}
$|\psi_0 \rangle$ (for even $N$ the
ground state is unique). We then perform a second NRG
run for the other Hamiltonian ${\cal H}^b$, during
which we construct the overlap matrices $S(m)$.
Technically these two NRG runs can be performed in 
parallel, in which case the overlap matrices $S(m)$
are calculated at the end of each NRG iteration.
Backtracking from iteration $N$ to iteration
$m_{\rm min}$, the reduced density matrices
$\rho^{\rm red \, (a)}(m; 0)$ are computed using
the standard TD-NRG algorithm.\cite{AndersSchiller2006}
These density matrices, along with the projected state
$|\chi^{\rm (a)}_M(0) \rangle$, are then used as
a seed for the hybrid-NRG algorithm detailed in
Eqs.~(\ref{A(t)-switching})--(\ref{eqn:chi-rho-evolution-2}).

\begin{figure}[tb]
\center{
\includegraphics[width=0.48\textwidth]{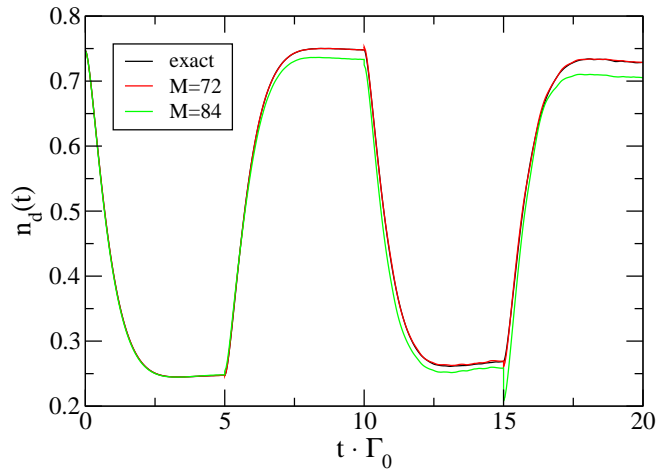}
}
\caption{(Color online)
         Comparison of the hybrid NRG-CET approach
         and exact diagonalization for the RLM on a
         Wilson chain with $\Lambda = 1.4$ and
         $N = 84$. The TD-NRG resummation is applied
         up to iteration $M$, beyond which the CET
         is used to track
         the time evolution of $|\chi_M(t) \rangle$.
         The number of states retained in the course
         of the NRG is equal to $N_s = 1024$. The
         resulting occupancy $n_d(t)$ is averaged over
         $N_z = 8$ equally distributed values of the
         twist parameter $z$, both for the hybrid-NRG
         and in exact diagonalization. Model
         parameters: $\Gamma_a = \Gamma_b = \Gamma_0$,
         $E_d^{b} = -E_d^{a} = \Gamma_0$, and
         $\tau = 5/\Gamma_0$, with $\Gamma_0/D = 10^{-5}$.
         Upon decreasing $M$, the hybrid NRG-CET
         approach gradually converges onto the exact
         result.
         }
\label{fig:switching5}
\end{figure}

\subsubsection{Simple extension of the TD-NRG to
               periodic switching}
               \label{sec:td-nrg-periodic-switching}

Before discussing the full hybrid approach, let us focus
for a moment on a simple extension of the TD-NRG to
periodic switching, corresponding to setting $M = N$ in
Eqs.~(\ref{A(t)-switching})--(\ref{eqn:chi-rho-evolution-2}).
This implies taking $|\chi_M(t) \rangle = 0$ throughout
the calculation.

Using $\tau = 5/\Gamma_0$ and
$E_d^{b} = -E_d^{a} = \Gamma_0$, Fig.~\ref{fig:switching5}
depicts a comparison of the TD-NRG with an exact
diagonalization solution on a Wilson chain with
$N = M = 84$. Since the decay rate is given by
$\Gamma_0$, the system almost fully equilibrates before
the next switching event takes place. As a result the
deviations of the periodic TD-NRG from the exact
numerical solution are surprisingly small. In particular,
the periodic TD-NRG correctly tracks the general
structure of the precise solution. Nevertheless, a
systematic degradation in accuracy is observed upon
going from one switching cycle to the next. This loss of
accuracy can be quantified by the growing discontinuity 
$\Delta n_d^{(i)} = | n_d(t_i + 0^+)- n_d(t_i - 0^+) |$
at successive switching events, caused by the
approximation made to $\rho^{\rm red \, (\alpha)}(m; t)$.

Reduction of the switching time to $\tau=1/\Gamma_0$
prevents the system from relaxing to a new equilibrium
state. In this case high-energy excitations, which are
cut off by neglecting the three additional terms
$\delta \rho^{+-}(m)$, $\delta \rho^{-+}(m)$, and
$\delta \rho^{--}(m)$, contribute significantly to the
time evolution of $\hat \rho(t)$. Since switching occurs
on a shorter time scale energy is frequently pumped
into the system, exciting it in such a way that the
three neglected high-energy contributions to
$\rho^{\rm red \, (\alpha)}(m; t)$ gain increasing
importance with time. Indeed, the periodic TD-NRG
result for $n_d(t)$, depicted by the blue curve in
Fig.~\ref{fig:switching1}(b), shows a sizable
accumulated error that grows systematically in time.

\subsubsection{Hybrid approach to periodic switching}
\label{sec:hybrid-periodic}

Although the simple periodic extension of the TD-NRG
already captures the correct short-time dynamics, the
externally driven nonequilibrium state increasingly
deviates with time from the transient dynamics of a single
quantum quench. Energy is locally added and removed from
the system, which can be partially dissipated via
heat current flowing between the impurity and the
finite-size bath.

To properly capture this physics, we next employ the
hybrid approach to periodic switching. The key idea
is to identify a suitable low-energy subspace that
is large enough for the neglected contributions in
our approximation to be small. A trivial limit
is given by setting $M \leq m_{\rm min}$, for which
the hybrid approach reproduces by construction the
exact solution on the finite-size chain. Even though
this limit has no practical value, it illustrates the
point that a reduction in $M$ should systematically
improve the quality of the results.

With this understanding, we extend our discussion
of the hybrid approach to $m_{\rm min} < M < N$.
At each time interval where ${\cal H}$ is fixed
we evolve $|\chi_M^{(\alpha)}(t) \rangle$ in time
using the CET, 
which is essentially exact on all
time scales of interest. The CET is restricted,
however, in the size of the Hilbert space one can
treat, which limits the length of the residual
chain $R_{M, N}$ extending beyond site $M$. With
our machines we can comfortably access up to
$2^{22} \approx 4\times 10^{6}$ basis states, above
which the computational effort rapidly becomes
too exhaustive. Keeping $N_s = 2^{10} = 1024$ states
at the conclusion of each NRG iteration, this sets
an upper limit of $12$ addition Wilson sites beyond
site $M$, i.e., $M \geq N - 12 = 72$. To reduce
finite-size effects we average over $N_z$ equally
distributed values of the twist
parameter~\cite{YoshidaWithakerOliveira1990}
$z \in (0, 1]$. Below we present results for
different values of $M \geq 72$ and $N_z \geq 1$.

\begin{figure}[tb]
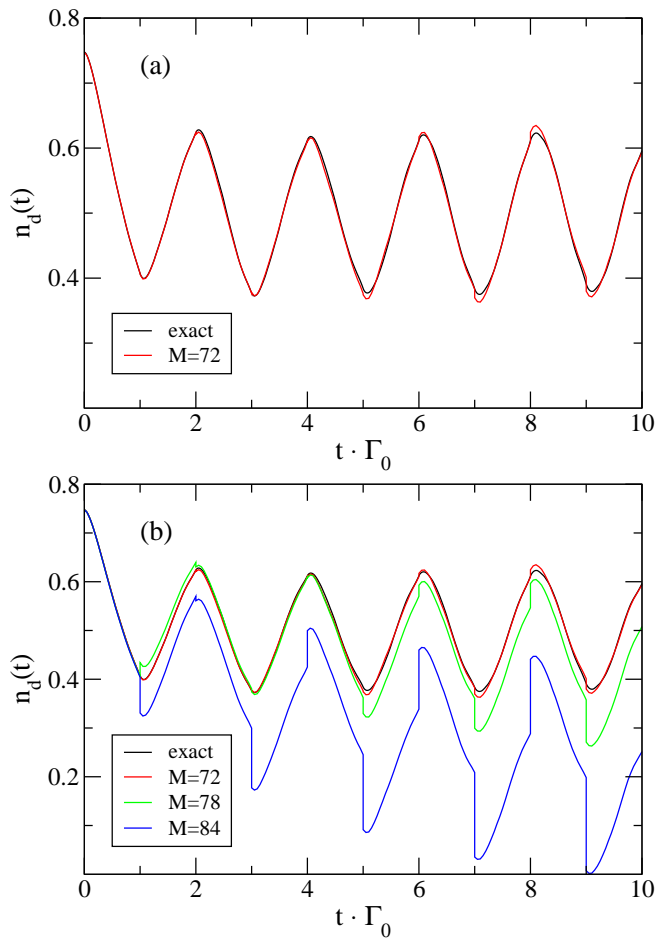

\begin{center}
 \includegraphics[width=0.48\textwidth]{fig7a.eps}
\includegraphics[width=0.48\textwidth]{fig7b.eps}
\end{center}
\caption{(Color online)
         Same as Fig.~\ref{fig:switching5} but for
         $\tau = 1/\Gamma_0$ and $N_z = 1$ (i.e.,
         no $z$ averaging). Note the much stronger
         $M$ dependence as compared to
         $\tau = 5/\Gamma_0$.}
\label{fig:switching1}
\end{figure}

Figures~\ref{fig:switching5} and \ref{fig:switching1}
show the time evolution of $n_d(t)$ in response to
repeated switchings between $E_d^a = -\Gamma_0$
and $E_d^b = \Gamma_0$, keeping
$\Gamma_a = \Gamma_b = \Gamma_0$ fixed.
Figure~\ref{fig:switching5} depicts four successive
switch events with the time interval $\tau = 5/\Gamma_0$,
while Fig.~\ref{fig:switching1} shows ten successive
switch events with the shorter time interval
$\tau = 1/\Gamma_0$. For comparison, the exact time
evolutions on the Wilson chain are depicted by the
black lines, after averaging over the different values
of $z$. These solutions are obtained by exact
diagonalization of the single-particle eigenmodes
of each Hamiltonian and using them to propagate
$d^{\dagger}(t) d(t)$ in time.

Several points are noteworthy. Upon decreasing $M$,
the hybrid NRG-CET curves gradually converge onto
the exact result for the Wilson chain, both for
$\tau = 1/\Gamma_0$ and $\tau = 5/\Gamma_0$.
Specifically, by $M = 72$ the hybrid NRG-CET approach
essentially reproduces the exact curves. The largest
deviations are usually found immediately following a
switching event, when a discontinuity generally develops
in the occupancy produced by the hybrid NRG-CET
approach. As noted above, the discontinuity stems
from the approximations employed in deriving
Eqs.~(\ref{chi_M-switching-1})--(\ref{eqn:chi-rho-evolution-2}).
It decreases in size with decreasing $M$, reflecting
a reduction in the accumulated weight of the
$|\phi_m(t)\rangle$'s in favor of $|\chi_M(t)\rangle$
in the expansion of Eq.~(\ref{partitioning-of-psi}).
Indeed, the smaller the accumulated weight of the
$|\phi_m(t)\rangle$'s the more accurate are the
approximations employed in deriving the hybrid-NRG.

For $M = 84$, the hybrid approach corresponds to the
simple extension of the TD-NRG to periodic switching.
As can be seen in Fig.~\ref{fig:switching1}(b), the
discontinuities at $t_n = n \tau$ are particularly
large for the shorter switching time $\tau = 1/\Gamma_0$ 
since more energy is pumped into the system. These
discontinuities are greatly reduced for the longer
time interval $\tau = 5/\Gamma_0$ as depicted in
Fig.~\ref{fig:switching5}. As discussed above, this
behavior is correlated with the fact that the occupancy
$n_d(t)$ for $\tau = 5/\Gamma_0$ has nearly decayed in
each time interval to its new equilibrium value
corresponding to the Hamiltonian ${\cal H}^{\alpha}$
in that time segment. In other terms, the state of
the system behaves as if it has effectively decayed
to the new ground state, leaving behind only weak
reminiscence of the transient behavior.

\begin{figure}[tb]
\vspace*{5pt}
\includegraphics[width=0.48\textwidth]{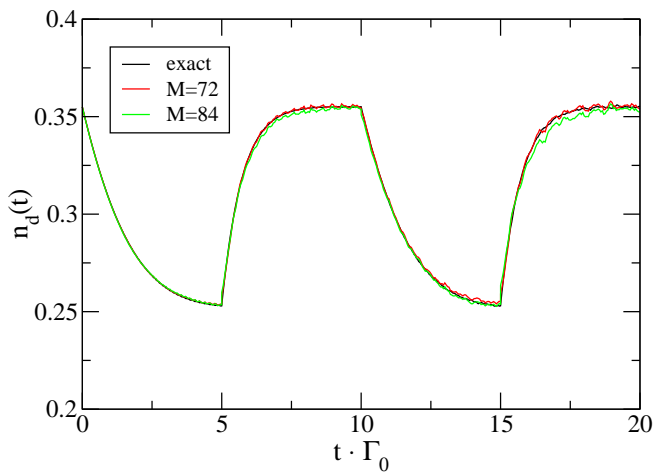}
\caption{(Color online)
         The time-dependent response to repeated
         switchings between $\Gamma_a = \Gamma_0/2$
         and $\Gamma_b = \Gamma_0$ for fixed
         $E_d^a = E_d^b = \Gamma_0/2$. All other
         parameters are the same as in
         Fig.~\ref{fig:switching5}.
         }
\label{fig:switching-of-Gamma}
\end{figure}

Figure~\ref{fig:switching-of-Gamma} shows the
complementary response to repeated switchings
between two different hybridization widths,
$\Gamma_a = \Gamma_0/2$ and $\Gamma_b = \Gamma_0$
for fixed $E_d^a = E_d^b = \Gamma_0/2$. As can be
seen, the discontinuities at $t_n = n \tau$ are
barely visible in this case, yet discretization
errors do give rise to high-frequency wiggles that are
essentially absent in Figs.~\ref{fig:switching5}
and \ref{fig:switching1}. The usage of
$z$-averaging is essential for reducing these
discretization effects, both at the level of
the hybrid-NRG and in the framework of exact
diagonalization. Similar to the case of a single
quantum quench, discretization errors become more
and more pronounced the larger are the deviations
between ${\cal H}^a$ and ${\cal H}^b$ and the longer
the time that has elapsed. Clearly, the nature
of the perturbation (e.g., $E_d^a \neq E_d^b$ vs
$\Gamma_a \neq \Gamma_b$) is also of importance.
It should be emphasized, however, that $z$-averaging 
alone is insufficient for removing low-frequency
oscillations that appear when the deviations between
${\cal H}^a$ and ${\cal H}^b$ are large.

To summarize this subsection, we have demonstrated
how the NRG can be used to systematically construct
effective low-energy Hamiltonians $\hat h_\chi^{a}$
and $\hat h_\chi^{b}$ whose bandwidth is smaller by
orders of magnitude than the bare conduction-electron
bandwidth $D$ of the original model. As a result
methods such as the TD-DMRG, whose accuracy is usually
confined to times of order $10^{2}/D$, can be used
within the hybrid platform to access exponentially long
time scales. This follows from the fact that the new
effective bandwidth $\approx D_M \propto D \Lambda ^{-M/2}$
can be exponentially smaller than $D$.

\subsection{Interacting resonant-level model}
\label{sec:IRLM}

Having established the accuracy of the hybrid NRG-CET
approach for the noninteracting RLM, we proceed to
apply it to an interacting problem that lacks an exact
reference solution. Specifically, we shall use the
hybrid NRG-CET to track the real-time dynamics of the
interacting resonant-level model (IRLM) in a regime not
accessible to other available methods. In the
IRLM,~\cite{Schlottmann1980,MethaAndrei2005,BoulatSaleurSchmitteckert2008,
BordaZawa2010} the resonant-level model of
Eq.~(\ref{H-NI-RLM}) is supplemented by a local
contact interaction $U$ between the level and the
band electrons:
\begin{eqnarray}
\H_U &=& U \left (
                    d^{\dagger} d - \frac{1}{2}
           \right )
           \frac{1}{N_k} \sum_{k, k'}
                  :\! c^{\dagger}_{k} c^{}_{k'} \!: . 
\label{eqn:IRLM-Hu}                  
\end{eqnarray}
Here $:\! c^{\dagger}_{k} c^{}_{k'} \!\!: = c^{\dagger}_{k}
c^{}_{k'} - \delta_{k, k'} \theta(-\epsilon_k)$ stands
for normal ordering with respect to the filled Fermi sea.
Physically, the contact interaction $U$ accounts for the
local capacitive coupling between the localized level
$d^{\dagger}$ and the band electrons. The total
Hamiltonian of the model is given by
\begin{eqnarray}
\H &=& \H_{RLM} + \H_U \; ,
\label{H-IRLM}
\end{eqnarray}
where $\H_{RLM}$ represents the RLM Hamiltonian of
Eq.~(\ref{H-NI-RLM}). 

At low energies, the IRLM is equivalent to a
renormalized noninteracting RLM, both describing a
phase-shifted Fermi liquid.\cite{Schlottmann1982,BordaSchillerZawadowski2008} 
At resonance the model
is characterized by the renormalized tunneling rate
\begin{equation}
\Gamma_{\rm eff}
           \approx D \left (
                              \frac{\Gamma_0}{D}
                     \right )^{1/(1 + \alpha)} ,
\end{equation}
where $\Gamma_0 = \pi \varrho V^2$ is the
noninteracting hybridization width of the RLM defined in
Eq.~(\ref{H-NI-RLM}), $\alpha$ equals
$2 \tilde{\delta} - \tilde{\delta}^2$, and
\begin{equation}
\tilde{\delta} = (2/\pi) \arctan(\pi \varrho U/2)
\end{equation}
is the scattering phase shift associated with $U$
alone (namely, in the absence of $V$). Within the
NRG, $\Gamma_{\rm eff}$ is conveniently extracted
from the zero-temperature charge susceptibility
$\chi_c = -d n_d/d E_d$, evaluated at $E_d = 0$.
Explicitly, we adopt the working definition
$\Gamma_{\rm eff} = 1/(\pi \chi_c)$.

\begin{figure}[tb]
\center{
\includegraphics[width=0.48\textwidth]{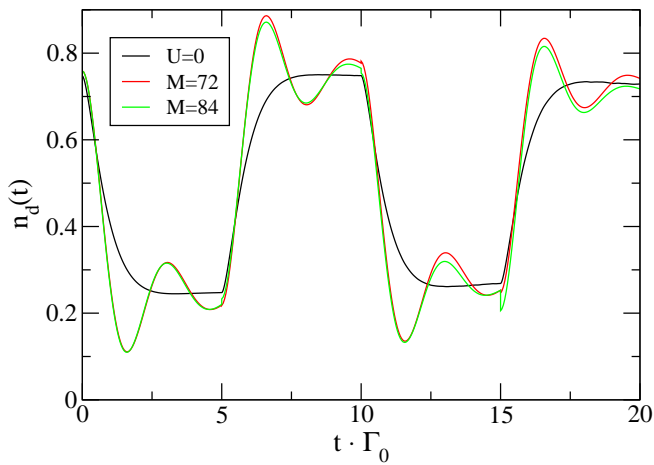}
}
\caption{(Color online)
         Same as Fig.~\ref{fig:switching5}, for $U/D = 1$
         and $E_d^b = -E_d^a = \Gamma_{\rm eff}$. Here
         $\Gamma_a = \Gamma_b = \Gamma_0 =
         6 \times 10^{-9}D$ was adjusted so as to maintain
         a fixed $\Gamma_{\rm eff} = 10^{-5} D$. The
         black line shows for comparison the exact time
         evolution on the Wilson chain for $U = 0$ and
         $\Gamma_a = \Gamma_b = \Gamma_0 = 10^{-5}D$,
         (taken from Fig.~\ref{fig:switching5}).
         }
\label{fig:switching-with-U}
\end{figure}

In Fig.~\ref{fig:switching-with-U} we show the time
evolution of $n_d(t)$ in response to repeated
switchings between $E_d^{a} = -\Gamma_{\rm eff}$ and
$E_d^b = \Gamma_{\rm eff}$ for $U/D = 1$. The bare
hybridization strength $\Gamma_a = \Gamma_b = \Gamma_0
= 6 \times 10^{-9} D$ was adjusted numerically so as
to maintain a fixed $\Gamma_{\rm eff} = 10^{-5} D$.
The effect of a finite $U$ is two-fold. First, it
renormalizes the bare hybridization width from
$\Gamma_0$ to $\Gamma_{\rm eff}$, which sets the
basic time scale in the problem: $1/\Gamma_{\rm eff}$.
Second, there are pronounced oscillations that are
absent for $U = 0$, having the characteristic
period $t_{\rm osc} \approx 3/\Gamma_{\rm eff}$. To
illustrate this point, we have borrowed from
Fig.~\ref{fig:switching5} the exact time evolution
of $n_d(t)$ on the Wilson chain for $U = 0$ and
$\Gamma_a = \Gamma_b = \Gamma_0 = 10^{-5}D$. The
effect of $U$ clearly goes beyond just a simple
renormalization of the parameters of the noninteracting
RLM, generating new interaction-induced oscillations.

\begin{figure}[tb]
\center{
\includegraphics[width=0.48\textwidth]{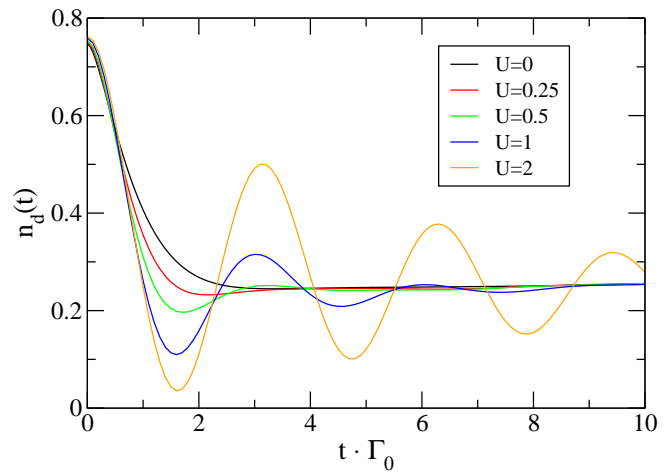}
}
\caption{(Color online)
         The real-time dynamics following a single
         quantum quench from $E_d^a = -\Gamma_{\rm eff}$
         to $E_d^b = \Gamma_{\rm eff}$, for the IRLM
         with $\Gamma_a = \Gamma_b = \Gamma_0$ and
         different values of the Coulomb repulsion $U$.
         Here $\Gamma_0$ was adjusted separately for
         each value of $U$ so as to maintain a fixed
         $\Gamma_{\rm eff} = 10^{-5}D$. All remaining
         parameters are as in Fig.~\ref{fig:switching5}.
         }
\label{fig:quench-with-U}
\end{figure}

To further investigate this point, we have plotted
in Fig.~\ref{fig:quench-with-U} the real-time dynamics
in response to a single quantum quench from $E_d^a =
-\Gamma_{\rm eff}$ to $E_d^b = \Gamma_{\rm eff}$,
for different values of the Coulomb repulsion $U$. The
bare hybridization width $\Gamma_a = \Gamma_b = \Gamma_0$
was adjusted separately for each value of $U$ so as
to maintain a fixed $\Gamma_{\rm eff} = 10^{-5}D$.
All curves begin at essentially the same initial
occupancy $n_d \simeq 0.75$ and decay at long time to
$n_d \simeq 0.25$. Hence the low-energy fixed point,
which governs the thermodynamics, is fully determined
by $\Gamma_{\rm eff}$ and $E_d$.

The intermediate dynamics, on the other hand, is quite
sensitive to $U$.  Starting from $U = 0$ and increasing
$U$, damped oscillations gradually develop with a
characteristic period that only weakly depends on $U$.
In contrast, the amplitude of the oscillations and their
damping rate strongly depend on $U$. The larger is $U$
the slower does the envelope function decays to zero,
resulting in the emergence of two distinct time scales
for large $U$: the period of oscillations $t_{\rm osc}$
and the characteristic damping time $t_{\rm damp}$. For
example, while $t_{\rm osc} \simeq 3.1/\Gamma_{\rm eff}$
for $U/D = 2$, the corresponding damping time is of order
$10/\Gamma_{\rm eff}$.

To understand the origin of these interaction-induced
oscillations, it is instructive to consider the limit
of a large Coulomb repulsion, $U \to \infty$. In this
extreme the level degree of freedom $d^{\dagger}$
and the zeroth Wilson shell $f^{\dagger}_0$ decouple from
the rest of the chain, being confined to a total valence
of one: $d^{\dagger}d + f^{\dagger}_0 f_0 = 1$ [this
valence is fixed by the normal-ordered form of
$\H_U$ as defined in Eq.~(\ref{eqn:IRLM-Hu})]. 
The corresponding subspace comprises of the two
configurations $d^{\dagger} |0\rangle$ and
$f_0^{\dagger} |0\rangle$, where $|0\rangle$ denotes the
state in which the resonant level and the zeroth Wilson
shell are both empty. Omitting the coupling to the
rest of the chain, the problem has been reduced in
effect to a $2 \times 2$ matrix whose eigenenergies are
\begin{equation}
\epsilon_{\pm} = \frac{E_d}{2}
          \pm \sqrt{
                     \left (
                             \frac{E_d}{2}
                     \right )^2
                     + V^2
                   } \; .
\end{equation}
Consequently, $n_d(t)$ must display quantum beats with
the frequency
\begin{equation}
\Omega = \epsilon_{+} - \epsilon_{-}
       = 2 \sqrt{
                  \left (
                          \frac{E_d}{2}
                  \right )^2
                  + V^2
                } \; .
\label{Omega-for-large-U}
\end{equation}

To make contact with Fig.~\ref{fig:quench-with-U},
it is necessary to express the period $t_{\rm osc}
= 2\pi/\Omega$ in terms of $\Gamma_{\rm eff} =
1/(\pi \chi_c)$. A straightforward calculation gives
\begin{equation}
n_d(E_d) = \frac{1}{2}
         - \frac{1}{4}
           \frac{E_d}{\sqrt{ (E_d/2)^2 + V^2}} \; ,
\end{equation}
resulting in $\Gamma_{\rm eff} = 4V/\pi$.
Substituting $V = (\pi/4) \Gamma_{\rm eff}$ into
Eq.~(\ref{Omega-for-large-U}) then yields
\begin{equation}
t_{\rm osc} =
   \frac{4}{\Gamma_{\rm eff}} \cdot
   \frac{1}
        {\sqrt{1 + (2 E_d/\pi \Gamma_{\rm eff})^2} } \; .
\end{equation}
Finally, setting $E_d = \Gamma_{\rm eff}$ to match the
value used in the curves of Fig.~\ref{fig:quench-with-U}
one obtains $t_{\rm osc} = 3.37/\Gamma_{\rm eff}$, in
excellent agreement with the period observed in
Fig.~\ref{fig:quench-with-U}.

Our strong-coupling analysis clearly reveals that
$U$ enters the quantum-beat frequency only via
$\Gamma_{\rm eff}$. However, $U$ strongly influences
the amplitude of the oscillations and their damping
rate. In contrast to the period of oscillations
which is well described by $U \to \infty$, the
damping time $t_{\rm damp}$ requires a finite
coupling to the rest of the chain. For $U\to \infty$
the impurity and the first Wilson shell decouple
from the rest of the chain, resulting in coherent
quantum beats between the two singly occupied
eigenstates. Once $U$ becomes finite the system
can decay incoherently to the lowest of the two
singly occupied states through a residual coupling
to the rest of the chain. It is this decay that
determines $t_{\rm damp}$. Since the residual coupling
should scale as $1/U$ for large interactions,
$t_{\rm damp}$ should scale as $U^2$ for $U \gg D$.
This corresponds to a damping rate that falls off
as $1/U^2$.

\section{Discussion and outlook}
\label{sec:discussion-outlook}

In this paper, we have extended the TD-NRG in two
ways. First, we devised a platform for combining
the TD-NRG with complementary methods such as the
TD-DMRG and CET for tracking the real-time dynamics
of quantum-impurity systems. Second, we extended
the TD-NRG from its original realm of a single
quantum quench to more complicated forms of
driven dynamics where repeated switchings are applied
to the system. As a proof of principle we combined
the TD-NRG with the CET to compute the response of
a resonant level, both with and without interactions,
to repeated switchings of its energy level and
tunneling amplitude to the band. Such a model can be
used to describe a single Coulomb-blockade resonance
in small quantum dots in regimes where spin degeneracy
is unimportant.

In the absence of interactions we have critically examined
our new approach by detailed comparisons to an exact
evaluation of the time evolution on the Wilson chain.
As long as the perturbations are not too large, good
accuracy is maintained over a fairly large number of
switching events. Usage of the hybrid NRG-CET greatly
improves the accuracy in cases where large deviations
develop between the exact curve and the one produced
without invoking the CET, see, e.g.,
Fig.~\ref{fig:switching1}. Our present implementation
of the hybrid approach is subject to the inherent
restriction of the CET to rather small finite-size
systems. We expect a significant boost in accuracy
and flexibility by combining the TD-NRG with the
TD-DMRG, which is capable of treating far larger
systems. In particular, our approach opens the
possibility to boost the TD-DMRG to exponentially
long time scales by using the TD-NRG to (i) construct
an effective low-energy Hamiltonian and (ii) account
for the short-time dynamics.
This portion of the research is left for
future work.

The potential of the new hybrid approach was next
demonstrated by applying it to repeated switchings
in the IRLM. Apart from specially tuned
models,~\cite{HeylKehrein2010} this constitutes to
our knowledge the first study of such driven
dynamics in interacting quantum impurity systems.
Although equivalent at sufficiently low energies
to a noninteracting RLM, the driven dynamics of
the IRLM shows clear traces of the interaction.
In particular, interaction-induced oscillations
develop for strong Coulomb repulsion, which have
no equivalent in the absence of interactions.
We were able to explain the period of oscillations
and its weak dependence on $U$ by invoking
the limit $U \to \infty$. The associated damping
time scales as $U^2$, and can greatly exceed the
natural damping time $1/\Gamma_0$ of the noninteracting
RLM. Such a long damping time is strictly the
effect of interactions.

\section*{Acknowledgments}
We are grateful to Dotan Goberman, Achim Rosch, Sebastian Schmitt
and Matthias Vojta  for helpful discussions.
This work was supported by the German-Israeli
Foundation through grant no.\ 1035-36.14, by
the Israel Science Foundation through grant
no.\ 1524/07 (E.E.\ and A.S.), and by the Deutsche
Forschungsgemeinschaft under AN 275/6-2
(F.G.\ and F.B.A).

%

\end{document}